\documentclass[10pt]{article}

\usepackage{amsmath}
\usepackage{amsfonts}
\usepackage{amssymb}

\usepackage[dvips]{graphicx}
\usepackage{subfigure}

\begin{document}

\newcommand{\be}{\begin{equation}}\newcommand{\ee}{\end{equation}}
\newcommand{\bea}{\begin{eqnarray}} \newcommand{\eea}{\end{eqnarray}}
\newcommand{\ba}[1]{\begin{array}{#1}} \newcommand{\ea}{\end{array}}

\numberwithin{equation}{section}

\def\np{Nucl. Phys. {\bf B}}\def\pl{Phys. Lett. {\bf B}}
\def\mpl{Mod. Phys. {\bf A}}\def\ijmp{Int. J. Mod. Phys. {\bf A}}
\def\cmp{Comm. Math. Phys.}\def\prd{Phys. Rev. {\bf D}}

\def\oa{\bigcirc\!\!\!\! a}
\def\ob{\bigcirc\!\!\!\! b}
\def\oc{\bigcirc\!\!\!\! c}
\def\oi{\bigcirc\!\!\!\! i}
\def\oj{\bigcirc\!\!\!\! j}
\def\ok{\bigcirc\!\!\!\! k}
\def\ve{\vec e}\def\vk{\vec k}\def\vn{\vec n}\def\vp{\vec p}
\def\vr{\vec r}\def\vs{\vec s}\def\vt{\vec t}\def\vu{\vec u}
\def\vv{\vec v}\def\vx{\vec x}\def\vy{\vec y}\def\vz{\vec z}

\def\ve{\vec e}\def\vk{\vec k}\def\vn{\vec n}\def\vp{\vec p}
\def\vr{\vec r}\def\vs{\vec s}\def\vt{\vec t}\def\vu{\vec u}
\def\vv{\vec v}\def\vx{\vec x}\def\vy{\vec y}\def\vz{\vec z}

\newcommand{\AdS}{\mathrm{AdS}}
\newcommand{\dd}{\mathrm{d}}
\newcommand{\eee}{\mathrm{e}}
\newcommand{\sgn}{\mathop{\mathrm{sgn}}}

\def\a{\alpha}
\def\b{\beta}
\def\g{\gamma}

\begin{flushright}
ICCUB-09-231\\
La Plata Th/09-01\\
September, 2009
\end{flushright}

\bigskip

\begin{center}

{\Large\bf  Tensionless supersymmetric  M2 branes in $\mathrm{AdS}_4 \times S^7$
\\
and Giant Diabolo}
\bigskip
\bigskip

{\it \large Jaume L{\'o}pez Carballo\footnotemark[1], Adri\'{a}n R. Lugo\footnotemark[2]\footnotemark[3], Jorge G. Russo\footnotemark[1]\footnotemark[4]}
\bigskip

{\it
1) Institute of Cosmos Sciences and Estructura i Constituents de la Materia\\
Facultat de F{\'\i}sica, Universitat de Barcelona\\
Barcelona, Spain \\
\smallskip
2) The Abdus Salam International Center for Theoretical Physics\\
Strada Costiera 11, (34014) Trieste, Italy\\ \smallskip
3) Departamento de F\'\i sica and IFLP-CONICET\\
Facultad de Ciencias Exactas, Universidad Nacional de La Plata\\
C.C. 67, (1900) La Plata, Argentina
\footnote{Permanent address.}
\\ \smallskip
4) Instituci\'o Catalana de Recerca i Estudis Avan\c cats (ICREA)\\
}
\bigskip
\bigskip

\end{center}
\bigskip

\begin{abstract}
We find various supersymmetric configurations of toroidal M2 brane solutions in $\AdS_4 \times S^7$ or, more generally,
in $\AdS_4 \times S^7/\mathbb{Z}_k$.
In this class we identify  solutions preserving   1/4 and 1/8 supersymmetries of the background.
The supersymmetric M2 branes have  angular momenta and winding on $S^7$, and null world-volumes.
In certain cases they collapse to string-like configurations.
These configurations can be viewed as a higher-dimensional (membrane) analog of BMN states.
We compute the energy and angular momenta, showing that all  supersymmetric configurations obey the BPS relation
$E=J/R $,\ $J\equiv \sum_{i=1}^4 |J_i|$ with $E, J\to \infty $.
Finally, we also study another class  of  supersymmetric M2-branes, including uncompact rotating membranes of  ``diabolo" shape.
\end{abstract}

\clearpage

\tableofcontents

\section{Introduction}

An important aspect of the AdS/CFT correspondence~\cite{malda,gubser,witten} is understanding
the precise map between supersymmetric states in the CFT and on the gravity side.
In the last years, there was an impressive  progress  in this direction.
In particular, the supergravity spectrum on $\AdS_5\times S^5$~\cite{romans} was put~\cite{witten} in precise correspondence with
the spectrum of 1/2 BPS operators of $N=4$ super Yang-Mills.

The correspondence between CFT operators and string states on AdS was generalized to various sectors in
\cite{BMN} for  (near BPS) collapsed string configurations and for more general extended string states in
numerous works (for reviews, see, e.g.~\cite{tseytlin}).
The correspondence between the spectra applies also to extended supersymmetric brane configurations,
such as BPS D brane configurations or giant gravitons, and
the identification of the corresponding operators led to important insights on the nature of the AdS/CFT correspondence~\cite{Hashimoto,bala,bere,LLM}.

The recent discovery of the ABJM superconformal field theory describing the physics of multiple membranes
probing an orbifold space~\cite{ABJM} provides an extremely interesting setup to
understand properties of AdS/CFT correspondence and of M-theory from a new perspective.
In a recent work~\cite{japos} BPS M2 brane configurations representing giant tori
were constructed.
The corresponding states carry a large amount of
angular momentum and D0 brane charge. The corresponding ABJM field theory interpretation was discussed in
\cite{berenstein}.
In this paper we will look for different types of supersymmetric configurations.
The general M2-brane solutions discussed here have a structure which is analog to that  of the
(non-supersymmetric) circular strings of~\cite{ART}.
These type of M2 brane solutions were investigated in~\cite{russo}. The configurations can also be viewed as
(toroidal) giant gravitons.
Here we will show that there is an important subclass of solutions which are supersymmetric
(general aspects of supersymmetric giant gravitons are discussed in~\cite{Mikhailov}).
This subclass of solutions has the property of having a vanishing determinant for the induced metric,
i.e. a null world-volume.
This is possible only for a tensionless membrane. They may be viewed as the large $J$ limit
of regular membranes. The solutions are the precise membrane analog of the tensionless strings discussed in~\cite{Mateos}.
They also represent  a higher dimensional version of the BMN states.

This  paper is organized as follows. In section 2 we review the $\AdS_4 \times S^7$ and $\AdS_4 \times S^7/\mathbb{Z}_k $ backgrounds
and their supersymmetries.
In section 3.1 we discuss the classical  equations of motion for an M2 brane moving in $\AdS_4 \times S^7$.
In section 3.2 we introduce our general ansatz that describes an M2 brane that rotates and winds in $S^7$,  and in section 3.3 we explicitly find
the values of winding number and angular velocities that solve all equations of motion.
In section 4 we derive  the BPS bound for the energy from the superalgebra.
In section 5 we find the energy formula for our membrane solutions   and show that in the supersymmetric limit
they reduce to the  expected  BPS form derived in section 4.
In section 6 we identify the subclass of solutions which preserve some fraction of supersymmetry.
Section 6.2 describes a class of regular supersymmetric membrane solutions, while section 6.3 discusses collapsed membrane configurations.
In section 7 the solutions are adapted to the case of $\AdS_4 \times S^7/\mathbb{Z}_k $ and, by dimensional reduction,
we obtain supersymmetric states in $\AdS_4 \times CP^3$.
In section 8 we revisit the giant torus rotating membrane solution found in \cite{japos} and show that in a certain region of the parameters
the rotating membrane opens up taking a ``diabolo" shape\footnote{The diabolo consists of a spool whirled and tossed on a string
(it illustrates  angular momentum conservation and it was said to be the favorite toy of Maxwell).}. We exhibit the solution in cylindrical coordinates,
where it has a simpler form, and present a convenient characterization of the torus, spiky membrane, diabolo, cylinder   and hyperboloid regimes in terms of a single parameter
(the last three solutions did not appear in \cite{japos}). In section 9 we present a summary of our results.
Appendix~\ref{appA} contains additional details of the calculations omitted
in the main text,  appendix B contains an alternative derivation of the supersymmetries of the collapsed membranes by
treating them as effective strings and in appendix C we give the expressions for the charges of the  solutions of section 8.

\section{Properties of $\AdS_4 \times S^7$ and $\AdS_4 \times S^7/\mathbb{Z}_k $ backgrounds}\label{sec:backgrounds}


The space $\AdS_4 \times S^7$ can be represented by the metric
\begin{equation}\label{metric}
\dd s^2 = \frac{R^2}{4} \big( \dd s^2_{\AdS_4} + 4\; \dd \Omega_7^2 \big) \ ,
\end{equation}
where $R = \ell_p (2^5 \pi^2 N)^{1/6}$, $\dd \Omega_7^2$ stands for the unit radius $S^7$ round metric, and
\begin{equation}
 \dd s^2_{\AdS_4} = - (1+r^2) \, \dd t^2 + \dfrac{\dd r^2}{1+r^2} + r^2 \big( \dd\theta^2 + \sin^2\theta\; \dd\varphi^2 \big) \ .
\end{equation}
The 4-form flux reads
\begin{equation}
 F^{(4)} = - \frac38 R^3\; r^2\, \sin\theta\; \dd t \wedge \dd r \wedge \dd\theta \wedge \dd\varphi \ .
\end{equation}

%
We can parametrize the $S^7$ using four complex coordinates, $Z^i$, which satisfy
\begin{equation}
 |Z^1|^2 + |Z^2|^2 + |Z^3|^2 + |Z^4|^2 = R^2 \ .
\end{equation}
Choosing
\begin{equation}\label{zi}
 Z^i = R\; \mu_i \;\eee^{i \xi^i}\ , \qquad \sum_{i=1}^4 \mu_i{}^2 = 1 \ ,
\end{equation}
the coordinates $\mu_i$ can be written in terms of hyper-spherical coordinates. A possible choice is
\begin{equation}
 \begin{aligned}
  \mu_1 & = \sin\alpha \ , \\
  \mu_2 & = \cos\alpha \sin\beta \ , \\
  \mu_3 & = \cos\alpha \cos\beta \sin\gamma \ , \\
  \mu_4 & = \cos\alpha \cos\beta \cos\gamma \ .
 \end{aligned}
\end{equation}
In these coordinates, the full metric reads
\begin{equation}\label{fullmetric}
 \begin{aligned}
  \dd s^2
   & = \frac{R^2}{4} \left\{ - (1+r^2) \, \dd t^2 + \dfrac{\dd r^2}{1+r^2} + r^2 \big( \dd\theta^2 + \sin^2\theta \;   \dd\varphi^2 \big)\right\}
\\ & \quad +
R^2 \left\{ \dd\alpha^2 + \cos^2\alpha\; \dd\beta^2 + \cos^2\alpha\; \cos^2\beta\; \dd\gamma^2 +
\sum_{i=1}^4 \mu_i{}^2\; \dd\xi^i{}^2 \right\}\ .
 \end{aligned}
\end{equation}

M-theory on $\AdS_4 \times S^7/\mathbb{Z}_k $ is obtained by identification under the $\mathbb{Z}_k$ orbifold action
\be
Z^i \to e^{i\frac{2\,\pi}{k}}\;Z^i\qquad\Longleftrightarrow\qquad\xi^i\to\xi^i+ \frac{2\pi}{k}\ ,
\label{orbi}
\ee
with integer $k$.
The solution represents the gravity dual of $ N$ M2-branes probing a ${\bf C}^4/\mathbb{Z}_k$ singularity, with $R$ equal to $\ell_p (2^5 \pi^2 Nk)^{1/6}$.
To connect with the ABJM theory it is useful to introduce $CP^3$ adapted variables. By completing squares we can write
\begin{equation}
 \dd \Omega_7^2 = \dd s^2_{CP^3} + (\dd y + A)^2 \ ,
\end{equation}
where $\dd A = 2 {\cal J}$ and ${\cal J}$ is the K\" ahler form of $CP^3$. We introduce a new set of coordinates adapted to $CP^3$, defined by
\begin{equation}\label{artan}
\begin{aligned}
\varphi_1 & = \xi^1 - \xi^2 \ , &
\varphi_2 & = \xi^3 - \xi^4 \ , \\
y & = \frac14 \big( \xi^1 + \xi^2 + \xi^3 + \xi^4 \big) \ , &
\psi & = \frac12 \big( \xi^1 + \xi^2 - \xi^3 - \xi^4 \big) \ , \\
\mu_1 & = \cos\zeta \cos\frac{\theta_1}{2} \ , &
\mu_2 & = \cos\zeta \sin\frac{\theta_1}{2} \ , \\
\mu_3 & = \sin\zeta \cos\frac{\theta_2}{2} \ , &
\mu_4 & = \sin\zeta \sin\frac{\theta_2}{2} \ .
\end{aligned}
\end{equation}
By reducing along $y$, we get type IIA strings on $\AdS_4 \times CP^3$,
\begin{subequations}
\begin{align}
\dd s^2 & = \tilde R^2 \big( \dd s^2_{\AdS_4} + 4\; \dd s^2_{CP^3} \big) \ , \qquad \tilde R^2 = \frac{1}{4k} R^3 \ , \\
\dd s^2_{CP^3} & = \dd \zeta^2 + \cos^2\zeta \,\sin^2\zeta\, \left( \dd\psi + \frac12\cos\theta_1\; \dd\varphi_1 -
\frac12 \cos\theta_2\; \dd\varphi_2 \right)^{\!\!2} \nonumber \\
& \qquad +\frac14 \cos^2 \zeta \,\Big( \dd\theta_1^2 + \sin^2\theta_1\; \dd\varphi_1^2 \Big)
+\frac14 \cos^2 \zeta\, \Big( \dd\theta_2^2 + \sin^2\theta_2 \;\dd\varphi_2^2 \Big) \ ,
\end{align}
\end{subequations}
with a one- and 3-form RR potentials and dilaton given by~\cite{Nilsson:1984bj}
\begin{align}
C^{(1)} & = \frac k2 \;\Big[ ( \cos^2\zeta - \sin^2\zeta )\;\dd\psi + \cos^2\zeta\, \cos\theta_1\; \dd\varphi_1
+ \sin^2\zeta\,\cos\theta_2\; \dd\varphi_2 \Big] = k\;A\ , \\
C^{(3)} & =  \frac{k}{2}\; \tilde R^2 \ r^3 \sin\theta\ \dd t \wedge \dd\theta \wedge \dd\varphi \ , \\
\eee^{2\phi} & = \frac{4\,\tilde R^2}{k^2}\ .
\end{align}


We now describe the supersymmetries of the background.
Our conventions for the Clifford algebras is such that $\{\Gamma_\mu,\ \Gamma_\nu\} = 2\, g_{\mu\nu}$,
where $g_{\mu\nu}$ is given by~\eqref{fullmetric}, and $\{\gamma_\mu,\ \gamma_\nu\} = 2\,\eta_{\mu\nu}$
is the standard flat space-time  Dirac algebra. We also define $\hat\gamma = - \gamma_{0123}$.
This allows us to write
\begin{equation}\label{gammas}
\begin{aligned}
\Gamma_t & = \frac R2\sqrt{1+r^2}\ \gamma_0 \ , &
\Gamma_r & = \frac R2 \frac1{\sqrt{1+r^2}}\ \gamma_1 \ , \\
\Gamma_\theta & = \frac R2\, r\ \gamma_2 \ , &
\Gamma_\varphi & = \frac R2\, r\sin\theta\ \gamma_3 \ , \\
\Gamma_\alpha & = R\ \gamma_4 \ , &
\Gamma_\beta & = R\cos\alpha\ \gamma_5 \ , \\
\Gamma_\gamma & = R\cos\alpha \cos\beta\ \gamma_6\ , &
\Gamma_{\xi^i} & = R\ \mu_i\ \gamma_{i+6}\ .
\end{aligned}
\end{equation}

The Killing spinors of this background are given by
\begin{equation}\label{susan}
\begin{aligned}
\epsilon &= \mathcal{M}\;\epsilon_0\ ,\\
\mathcal M &\equiv M_\alpha M_\beta M_\gamma \left( \prod_{i=1}^4 M_i \right) M_r M_t M_\theta M_\varphi\ .
\end{aligned}
\end{equation}
Here $\epsilon_0$ is an arbitrary constant Majorana spinor, and the $M_\mu$'s are the exponentiation of generators
of translations in the $\mu$-direction,
\begin{equation}
\begin{aligned}
M_t & = \eee^{\frac t 2 \hat\gamma \gamma_0} \ , &
M_r & = \eee^{\frac {\bar r} 2 \hat\gamma \gamma_1} \ , &
M_\theta & = \eee^{\frac \theta 2 \gamma_{12}} \ , &
M_\varphi & = \eee^{\frac \varphi 2 \gamma_{23}} \ , \\
M_\alpha & = \eee^{\frac \alpha 2 \hat\gamma \gamma_4} \ , &
M_\beta  & = \eee^{\frac \beta  2 \hat\gamma \gamma_5} \ , &
M_\gamma & = \eee^{\frac \gamma 2 \hat\gamma \gamma_6} \ , &
M_i & = \eee^{\frac {\xi^i} 2\mathbb X_i}\ .
\end{aligned}
\end{equation}
where we have defined $r = \sinh\hat r$, and introduced%
\footnote{
The last relation in (\ref{matX}) follows from the definition $\;\gamma_{10} \equiv -\gamma_0\,\gamma_1\dots\gamma_9$.
}
\be
(\mathbb X_i) \equiv (\gamma_{47},\ \gamma_{58},\ \gamma_{69},\ \hat\gamma\gamma_{10})\ , \qquad
\mathbb X_1\;\mathbb X_2\;\mathbb X_3\;\mathbb X_4 = -1 .
\label{matX}
\ee

Next, consider the $\mathbb{Z}_k$ orbifold action~(\ref{orbi}), which only affects to the $\xi^i$ angular variables.
%
%
Let us define the eigenvalues of $\mathbb X_1$, $\mathbb X_2$ and $\mathbb X_3$ to be $i \varsigma_i$.
Since $\mathbb X_i^2 = -1$, it must be $\varsigma_i = \pm 1$.
The spinors in (\ref{susan}) with $\varsigma_1 = \varsigma_2 = \varsigma_3$ are projected out by the
projection~(\ref{orbi}) with $k > 2$, henceforth 24 Killing spinors (3/4 of the original 32) survive the orbifold action.

\section{A class of M2 brane configurations}

\subsection{Action and equations of motion}

Let $Y^\mu$, with $\mu = 0, \cdots, 4$, be the embedding coordinates in the AdS piece of the space, and
$X^k$, $k = 1,\cdots, 8$, the ones corresponding to the 7-sphere. The  membrane action reads~\cite{russo}
\begin{equation}
\begin{aligned}
S & = \frac{T_2}{2} \int\!\! \dd^3\sigma\; \bigg( - \sqrt{-h}\; h^{\alpha\beta}
\big( \eta_{\mu\nu}\;\partial_\alpha Y^\mu \partial_\beta Y^\nu  +
\delta_{kj}\;\partial_\alpha X^k \partial_\beta X^j \big) +\sqrt{-h}\\
& + \tilde\Lambda \;\left( \eta_{\mu\nu}\; Y^\mu Y^\nu  +  \frac{R^2}{4} \right)
+ \Lambda \;( X^k X^k - R^2 ) \bigg) + T_2\; \int\; C^{(3)}|_{pullback}\ .
\end{aligned}
\end{equation}
%
We choose $\eta_{\mu\nu} = \text{diag} ( -1, 1, 1, 1, -1)$. $\tilde\Lambda$ and $\Lambda$ are Lagrange multipliers
that enforce the conditions
\begin{equation}
\eta_{\mu\nu}\;Y^\mu Y^\nu = - \frac{R^2}{4} \ , \qquad  \sum_{k=1}^4 (X^k)^2 = R^2 \ ,\label{cons}
\end{equation}
respectively, thus defining the $\AdS_4 \times S^7$ space.

Using the formula $\delta h = - h h_{\alpha\beta} \delta h^{\alpha\beta}$ one finds that the equation of motion for the world-volume metric gives
\begin{equation}
h_{\alpha\beta} = \eta_{\mu\nu}\; \partial_\alpha Y^\mu \partial_\beta Y^\nu  + \delta_{kj}\;\partial_\alpha X^k \partial_\beta X^j \ .
\end{equation}
The equations of motion for $Y^\mu$ and $X^k$ are given by
\begin{subequations}\label{eom}
\begin{align}
\partial_\beta \big( \sqrt{-h} h^{\alpha\beta} \partial_\alpha Y_\mu \big) & = -\tilde\Lambda\; Y_\mu\ , \\
\partial_\beta \big( \sqrt{-h} h^{\alpha\beta} \partial_\alpha X_k \big) & = - \Lambda\; X_k\ ,
\end{align}
\end{subequations}
where the indexes of $Y^\mu$ are lowered and raised by $\eta_{\mu\nu}$.
It is also useful to define the variables
\begin{subequations}
\begin{align}
Z_0 & = Y^0 + i Y^4 = \frac{R}{2}\;\sqrt{1+r^2}\; \eee^{it}\ ,\\
Y^i & =\frac{R}{2}\; r\;n^i\ , \qquad i= 1,2,3\ ,\\
Z^i& = X^{2i-1} + i\, X^{2i} = R\; \mu_i\; \eee^{i \xi^i} \ , \qquad i= 1,\dots, 4\ ,
\end{align}
\end{subequations}
where the constraints~(\ref{cons}) enforce $\vec n\cdot\vec n =1\;$ and $\sum\limits_{i=1}^4\mu_i{}^2 =1$;
their equations of motion read,
\begin{subequations}\label{eoms}
\begin{align}
\partial_\beta \big( \sqrt{-h} h^{\alpha\beta} \partial_\alpha Z_0 \big)  &= - \tilde\Lambda\; Z_0 \ , \\
\partial_\beta \big( \sqrt{-h} h^{\alpha\beta} \partial_\alpha \vec Y \big) & = - \tilde\Lambda\; \vec Y \ , \\
\partial_\beta \big( \sqrt{-h} h^{\alpha\beta} \partial_\alpha Z^i \big)  &= - \Lambda\; Z^i \ .
\end{align}
\end{subequations}

\subsection{General ansatz}

We now introduce the following ansatz,
\begin{equation}\label{ansatz}
\begin{aligned}
t & = \omega_0\; \sigma^0\ , \quad & r & = 0 \ , \\
\mu_i & = \text{constant} \ ,\qquad &
\xi^i & = \omega_i\; \sigma^0 + m_i\; \sigma^1 + n_i\; \sigma^2 \equiv \frac12 \beta_\alpha^{i} \sigma^\alpha\ ,
\end{aligned}
\end{equation}
where $\sigma^1 ,\ \sigma^2$ are $2\pi$-periodic.
Since $m_i$ and $n_i$ represent winding numbers, all of them must be integers; furthermore,
for convenience we have introduced the compact notation,
$\beta_0^i = 2\, \omega_i \;,\;\beta_1^i  = 2\; m_i \;,\; \beta_2^i= 2\; n_i$.\footnote{The index
$i$ in  $\omega_i,\ m_i,\ n_i$ has been written as a {\it subindex} to avoid confusion with powers in the formulas containing
specific values of $i$ (e.g. we prefer to write $m_2$ instead of $m^2$).}

Solutions with this structure were found in~\cite{russo} in a particular gauge where $h_{01}=h_{02}=0$, $h_{00}={\rm const.}(h_{12}^2-h_{11}h_{22})$.\footnote{Generalizations of the solutions of~\cite{russo} including non-constant $\mu_i$ were discussed in~\cite{Bozhilov} (extending the
integrable string $\sigma $ models of~\cite{ART} to membranes).}
However, we will be later interested in a special class of solutions (called ``non-collapsed membranes") for which this gauge choice is inconvenient.
Therefore the analysis of solutions will be carried out
in an arbitrary gauge.

The $i$ index of $\beta_\alpha^{\,i}$ can be raised with the $\xi^i$ part of the metric~(\ref{fullmetric}), i.e.,
\begin{equation}\label{compactansatz}
\beta_{i,\alpha} \equiv \mu_i{}^2\;\beta^i_\alpha \ , \qquad
\beta_{i,\alpha} \beta_\beta^{i}  \equiv \sum_{i=1}^4 \mu_i^2 \beta_\alpha^{i} \beta_\beta^{i}\ .
\end{equation}
The world-volume metric becomes
\begin{align}
h_{\alpha\beta} & = \frac{R^2}{4} \left( \beta_{i ,\alpha} \beta_\beta^{i} - \omega_0^2\; \delta_{\alpha,0}\delta_{\beta,0} \right) \ ,\label{hab} \\
h & = - \frac{R^6}{64} \Big\{ \omega_0^2 \Big[ \big( \beta_{i,1} \beta_1^{i} \big)  \big( \beta_{j,2} \beta_2^{j} \big) - \big(\beta_{i,1} \beta_2^{i}\big)^2 \Big] - \det\nolimits_{\alpha, \beta }\big( \beta_{i,\alpha} \beta_\beta^{i} \big) \Big\} \ . \label{h}
\end{align}
%

The ansatz~(\ref{ansatz}) includes momentum and winding around all four $\xi^i$ angles. However, by performing a redefinition
in the world-volume coordinates, we can  reduce it to a problem with rotation in two planes only. Namely, defining
\begin{equation}
\begin{aligned}
\tilde \sigma^0 & = \sigma^0 \ , \\
\tilde \sigma^1 & = \frac1{2} \big( \beta_0^{1}\; \sigma^0 + \beta_1^{1}\; \sigma^1 + \beta_2^{1}\; \sigma^2 \big)\ , \\
\tilde \sigma^2 & = \frac1{2} \big( \beta_0^{3}\; \sigma^0 + \beta_1^{3}\; \sigma^1 + \beta_2^{3}\; \sigma^2 \big)\ ,
\label{garin}
\end{aligned}
\end{equation}
the ansatz~(\ref{ansatz}) reduces to
\begin{equation}\label{2planesansatz}
\begin{aligned}
\xi^1 & = \tilde\sigma^1 \ ,&\qquad
\xi^2 & = \tilde \omega_2 \;\tilde\sigma^0 + \tilde m\; \tilde\sigma^1 + \tilde n'\; \tilde\sigma^2\ , \\
\xi^3 & = \tilde\sigma^2 \  ,&\qquad
\xi^4 & = \tilde \omega_4\; \tilde\sigma^0 +\tilde  m'\; \tilde\sigma^1 +\tilde  n\; \tilde\sigma^2 \ .
\end{aligned}
\end{equation}
It should  be noted that it is (locally) equivalent to the original~(\ref{ansatz}) only if the following condition holds,
\be
\beta_1^{1}\; \beta_2^{3} - \beta_2^{1}\; \beta_1^{3}\neq 0 \ .
\label{gro}
\ee
Because of the periodicity of the $\sigma^1,\ \sigma^2 $ variables,  the solutions are not globally  equivalent in general.
We recall that winding numbers must be integers for membranes in $AdS_4\times S^7$ (and $\in \mathbb{Z}/k$ for membranes in $AdS_4\times S^7/\mathbb{Z}_k$).

We will be interested in the particular case $\tilde m'=\tilde n'=0$,
i.e. in the solution
\begin{equation}
\begin{aligned}
\xi^1 & = \tilde\sigma^1 \  ,&\qquad
\xi^2 & = \tilde \omega_2 \;\tilde\sigma^0 + \tilde m\; \tilde\sigma^1 \ , \\
\xi^3 & = \tilde\sigma^2 \  ,&\qquad
\xi^4 & = \tilde \omega_4\; \tilde\sigma^0 + \tilde n\; \tilde\sigma^2 \ .
\label{azzar}
\end{aligned}
\end{equation}
Returning to the $\sigma^\alpha $ variables, (\ref{azzar}) corresponds to the following choice in eq.~(\ref{ansatz}),
\begin{equation}
\begin{aligned}
\vec {\omega} & = ( \omega_1 ,\ \omega_2,\ \omega_3 ,\ \omega_4 ) \ , \\
\vec {m} & = ( a ,\ \alpha\,a ,\ b ,\ \beta\, b ) \ , \\
\vec {n} & = ( c ,\ \alpha\, c ,\ d ,\ \beta\, d ) \ .
\label{abo}
\end{aligned}
\end{equation}
if we make the identifications,
\be
\tilde m \equiv \alpha\ ,\qquad\,\qquad \tilde n \equiv \beta\ ,\qquad\,\qquad
\tilde\omega_2\equiv \omega_2 -\alpha \,\omega_1 \ , \qquad\,\qquad \tilde\omega_4\equiv \omega_4 -\beta\,\omega_3\ .
\ee
Equations (\ref{garin}) then take the form
\begin{equation}
\begin{aligned}
\tilde\sigma^0 &\equiv \sigma^0 \ , \cr
\tilde\sigma^1 &\equiv \omega_1\;\sigma^0 + a\;\sigma^1 +c\;\sigma^2\ ,\cr
\tilde\sigma^2 &\equiv \omega_3\;\sigma^0 + b\;\sigma^1 +d\;\sigma^2\ ,
\label{garin2}
\end{aligned}
\end{equation}
and the condition (\ref{gro}) for this equivalence to hold now reads $ad-bc\neq 0$.
One has the option of considering $\tilde m,\ \tilde n$ integers in~(\ref{azzar}), or the solution~(\ref{ansatz}), (\ref{abo}), with $\vec m,\ \vec n$ integers,
giving rise to globally inequivalent solutions.

 The ansatz (\ref{azzar}) leads to the following values for $\beta^i_\alpha $:
\begin{align}
(\beta^1_\alpha) &= (0,\; 2,\; 0)\ ,&(\beta^2_\alpha)&= (2\,\tilde \omega_2,\; 2\,\tilde m,\; 0)\ ,&
(\beta^3_\alpha) &= (0,\; 0,\; 2)\ ,&(\beta^4_\alpha)&= (2\,\tilde \omega_4,\; 0,\; 2\,\tilde n)\ .
\label{betita}
\end{align}

When $a\,d-b\,c=0$, $\vec m$ results proportional to $\vec n$.
More generally, whenever $\vec m=K\vec n$, we have $\beta_1^{1}\; \beta_2^{3} - \beta_2^{1}\; \beta_1^{3}= 0 $
(or $ad-bc=0$) and the change of coordinates (\ref{garin}) (or (\ref{garin2})) is not possible.
Instead, it will be more convenient to  introduce a new world-volume coordinate
$\sigma =\sigma^2+K\,\sigma^1$, exhibiting the fact that the configuration depends only on $\sigma$.
This is the case when the  M2 brane collapses to a string-like configuration.

\subsection{Solving the conditions on the parameters}

The equations of motion~(\ref{eoms}) impose some conditions on the parameters characterizing the solution.
In order to solve these conditions for the  ansatz (\ref{azzar}),
we first compute the inverse matrix $h^{\alpha\beta} = \frac{h_c^{\alpha\beta}}{h}\ ,$
where $h_c^{\alpha\beta}$ is the co-factor matrix of $h_{\alpha\beta}$.
Its explicit expression is given in the appendix~\ref{appA} (for clarity in the notation, in this section and in the appendix
we will remove ``tildes" from $\tilde \omega_2,\ ,\tilde \omega_4,\ \tilde m,\ \tilde n$).

The equations of motion~(\ref{eoms}) then reduce to
%
\begin{align}
- \omega_0{}^2 \; h^{00}_c &= \sqrt{-h}\; \tilde\Lambda\label{Y} \ ,\\
-\frac{1}{4}\; h_c^{\alpha\beta}\;\beta^i_\alpha\;\beta^i_\beta &= \sqrt{-h}\;\Lambda
\ , \qquad i = 1,\dots,4\ . \label{XY}
\end{align}
%
While the first equation just fixes the value of $\tilde\Lambda$,
the second one gives non-trivial conditions, since it must be satisfied for each $i=1,...,4$.
One of the equations determines $\Lambda$ and, generically,  three independent conditions remain.

Using the expressions for $h^{\alpha\beta}_c$ given in the appendix~\ref{appA}, equations ~(\ref{Y})-(\ref{XY}) become
\be
-\frac{\,\sqrt{-h}}{\omega_0{}^2\;R_{}^ 4}\;\tilde\Lambda  =
(\mu_1{}^2 +\mu_2{}^2\;m^2)\;(\mu_3{}^2 +\mu_4{}^2\;n^2)\label{Y1} \ ,
\ee
\begin{subequations}\label{X1}
\begin{align}
-\frac{4\,\sqrt{-h}}{\omega_0^2\;R^4}\;\Lambda&=
(\mu_3{}^2+n^2\;\mu_4{}^2)\;\left(\mu_2{}^2\;\left(\frac{2\,\omega_2}{\omega_0}\right)^2 -1\right)
+ \mu_3{}^2\;\mu_4{}^2\; \left(\frac{2\,\omega_4}{\omega_0}\right)^{\!2} ,\\
-\frac{4\,\sqrt{-h}}{\omega_0^2\;R^4}\;\Lambda&=
(\mu_3{}^2+n^2\;\mu_4{}^2)\;\left(\mu_1{}^2\;\left(\frac{2\,\omega_2}{\omega_0}\right)^2 -m^2\right)
+ \mu_3{}^2\;\mu_4{}^2\; m^2\;\left(\frac{2\,\omega_4}{\omega_0}\right)^{\!2} ,\\
-\frac{4\,\sqrt{-h}}{\omega_0^2\;R^4}\;\Lambda&=
(\mu_1{}^2+m^2\;\mu_2{}^2)\;\left(\mu_4{}^2\;\left(\frac{2\,\omega_4}{\omega_0}\right)^2 -1\right)
+ \mu_1{}^2\;\mu_2{}^2\; \left(\frac{2\,\omega_2}{\omega_0}\right)^{\!2} ,\\
-\frac{4\,\sqrt{-h}}{\omega_0^2\;R^4}\;\Lambda&=
(\mu_1{}^2+m^2\;\mu_2{}^2)\;\left(\mu_3{}^2\;\left(\frac{2\,\omega_4}{\omega_0}\right)^2 -n^2\right)
+ \mu_1{}^2\;\mu_2{}^2\; n^2\;\left(\frac{2\,\omega_2}{\omega_0}\right)^{\!2} ,
\end{align}
\end{subequations}
where the determinant $h$ of $h_{\alpha\beta}$ is given by,
\begin{equation}\label{det}
\begin{aligned}
-\frac{4}{\omega_0{}^2\,R^6}\;h & =
(\mu_1{}^2+m^2\;\mu_2{}^2)\;(\mu_3{}^2+n^2\;\mu_4{}^2) -
\mu_1{}^2\;\mu_2{}^2\;(\mu_3{}^2+n^2\;\mu_4{}^2)\left(\frac{2\,\omega_2}{\omega_0}\right)^{\!2}
\\ & \quad
-\mu_3{}^2\;\mu_4{}^2\;(\mu_1{}^2+m^2\;\mu_2{}^2)\left(\frac{2\,\omega_4}{\omega_0}\right)^{\!2}\ .
\end{aligned}
\end{equation}
For generic%
\footnote{
By generic we mean that all the $\mu_i$'s are non zero and different from each other.
}%
values of the $\mu_i$'s
the relations~(\ref{X1}) impose three conditions on the parameters. They can be solved explicitly
in terms of a free variable $z$ as follows,
\begin{subequations}\label{sngral}
\begin{align}
m^2 &= \frac{\mu_1{}^2}{\mu_2{}^2}\;\frac{z- \mu_1{}^2 }{z - \mu_2{}^2}\ , \qquad\quad
n^2 = \frac{\mu_3{}^2}{\mu_4{}^2}\;\frac{z- \mu_3{}^2 }{z - \mu_4{}^2}\ ,\\
\left(\frac{2\,\omega_2}{\omega_0}\right)^2&=
\frac{1}{\mu_2{}^2}\,\frac{2\,z-\mu_1{}^2-\mu_2{}^2}{z-\mu_2{}^2}\frac{1}{(3\,z-1)(z-z_0)}\;
\left(z^2 - z_2\;z + (\mu_1{}^2+\mu_2{}^2)\,z_0\right)\ ,\\
\left(\frac{2\,\omega_4}{\omega_0}\right)^2&=
\frac{1}{\mu_4{}^2}\,\frac{2\,z-\mu_3{}^2-\mu_4{}^2}{z-\mu_4{}^2}\frac{1}{(3\,z-1)(z-z_0)}\;
\left(z^2- z_4\,z + (\mu_3{}^2+\mu_4{}^2)\,z_0\right)\ ,
\end{align}
\end{subequations}
where we have defined,
\begin{equation}
\begin{aligned}
z_0 &\equiv C_0 \Big( \mu_1{}^2\;\mu_4{}^2\,(\mu_2{}^2+\mu_3{}^2) - \mu_2{}^2\;\mu_3{}^2\,(\mu_1{}^2+\mu_4{}^2)\Big) \ ,  \\
z_2 &\equiv C_0 \Big( \mu_1{}^2\;\mu_4{}^2\,(1+\mu_2{}^2-\mu_4{}^2) - \mu_2{}^2\;\mu_3{}^2\,(1+\mu_1{}^2-\mu_3{}^2)\Big) \ , \\
z_4 &\equiv C_0 \Big( \mu_1{}^2\;\mu_4{}^2\,(1+\mu_3{}^2-\mu_1{}^2) - \mu_2{}^2\;\mu_3{}^2\,(1+\mu_4{}^2-\mu_2{}^2)\Big) \ , \\
\end{aligned}
\end{equation}
and $C_0  \equiv (\mu_1{}^2\;\mu_4{}^2 - \mu_2{}^2\;\mu_3{}^2)^{-1}$. They satisfy the relations,
\begin{align}
z_2 +z_4 -2\,z_0 & = 1\ , & z_2 - z_4 & = \mu_1{}^2 +\mu_2{}^2 - \mu_3{}^2 - \mu_4{}^2 \ .
\end{align}
For completeness, we also give  the expression for the Lagrange multiplier parameters,
\begin{equation}
\begin{aligned}
-\frac{\sqrt{-h}}{\omega_0{}^2\;R^4}\;\tilde\Lambda &=
\mu_1{}^2\;\mu_3{}^2\; \frac{2\,z - \mu_1{}^2 -\mu_2{}^2}{z-\mu_2{}^2}\;
\frac{2\,z - \mu_3{}^2 -\mu_4{}^2}{z-\mu_4{}^2}\ ,\\
-\frac{4\,\sqrt{-h}}{\omega_0{}^2\;R^4}\;\Lambda &= -\frac{1}{3\,z-1}\;
\mu_1{}^2\;\mu_3{}^2\; \frac{2\,z - \mu_1{}^2 -\mu_2{}^2}{z-\mu_2{}^2}\;
\frac{2\,z - \mu_3{}^2 -\mu_4{}^2}{z-\mu_4{}^2}\ .
\end{aligned}
\end{equation}
We find that the on-shell value of the determinant of the metric~(\ref{det})  is given by
\be
-\frac{4}{\omega_0{}^2\,R^6}\;h = \mu_1{}^2\;\mu_3{}^2\;
\frac{2\,z - \mu_1{}^2 -\mu_2{}^2}{z-\mu_2{}^2}\;
\frac{2\,z - \mu_3{}^2 -\mu_4{}^2}{z-\mu_4{}^2}\;\frac{z}{3\,z-1}\ .
\ee

Of particular interest is the case $z=0$, because it characterizes a supersymmetric solution (see section 6.2).
In this $h=0$ case the membrane becomes tensionless. A similar phenomenon for supersymmetric rotating strings had
been noticed  in  ~\cite{Mateos}. These M2 brane configurations with $h=0$ are thus the precise higher dimensional
analog of the rotating strings of \cite{Mateos}. When $z=0$ the winding numbers and angular velocities are (up to signs)
determined by the $\mu_i$ by  the following relations:
\begin{align}
m^2 & =\frac{\mu_1{}^4}{\mu_2{}^4} \ , &
n^2 & =  \frac{\mu_3{}^4}{\mu_4{}^4} \ , &
\left(\frac{2\,\omega_2}{\omega_0}\right)^{\!2} & = \left(1+\frac{\mu_1{}^2}{\mu_2{}^2}\right)^{\!2} , &
\left(\frac{2\,\omega_4}{\omega_0}\right)^{\!2} & = \left(1+\frac{\mu_3{}^2}{\mu_4{}^2}\right)^{\!2} ,
\label{snz=0}
\end{align}
where we have used equations (\ref{sngral}).
This solution is continuously connected with the $|z|=\infty$ solution, for which $h\neq 0$ and
\begin{align}
m^2& =\frac{\mu_1{}^2}{\mu_2{}^2}  \ , &
n^2 & =\frac{\mu_3{}^2}{\mu_4{}^2} \ , &
\left(\frac{2\,\omega_2}{\omega_0}\right)^2 & = \frac{2}{3\,\mu_2{}^2} \ , &
\left(\frac{2\,\omega_4}{\omega_0}\right)^2 & = \frac{2}{3\,\mu_4{}^2}\ .
\label{snz=infty}
\end{align}

\section{BPS bound from the superalgebra}

In this section we use the superalgebra on the  $\AdS_4 \times S^7$ vacuum to show  that  a solution that preserves
a fraction of the supersymmetries must obey a simple bound.
Our discussion follows the similar derivation given in~\cite{Mateos} for $\AdS_5\times S^5$.
The  $\AdS_4 \times S^7$ vacuum has the isometry superalgebra $OSp(4|8) $. The bosonic symmetry is
$ SO(2, 3) \times SO(8)$. The supercharges are 32 Majorana spinors which under the $ SO(2, 3) \times SO(8)$ subgroup
of the 11d Lorentz group $SO(1,10)$ decompose as 4-component Majorana spinors $Q_a$, with $a=1,...,8$, transforming in the spinorial ${\bf 8_s}$
representation of $SO(8)$ (more precisely, $spin(8)$). Let us denote by $\tilde \gamma_\mu $ (in this section $\mu,\nu=0,1,2,3$)  the $4\times 4$
four-dimensional Dirac matrices for $\AdS_4$.
The anticommutators are
\begin{equation}
\big\{ Q_a ,\ Q_b\big\} = C \left[ \left( \tilde \gamma_\mu \;P^\mu + \frac12\, \tilde \gamma_{\mu\nu}\; M^ {\mu\nu} \right)
\delta_{ab} + \mathbb{I}\; \hat B_{ab} \right]\ ,
\end{equation}
where $C$ is the charge conjugation matrix ($C = \tilde \gamma^0$ for the real Majorana representation), $P^\mu$, $M^{\mu\nu}$
are the charges in $\AdS_4$ , and $\hat B_{ab}$ is a real antisymmetric matrix of $spin(8)$  charges.
For our solutions, the only non-vanishing charges are the energy $P^0$ and the angular momenta $J_1,\dots,J_4$.
These last ones are eigenvalues of the Cartan generators  of $SO(8)$ in the vector representation ${\bf 8_v}$.
Using the standard relation $\hat B_{ab}={1\over 4} \hat \gamma_{ab}^{ij} B_{ij}$, where
$\{\hat \gamma^i, i=1,\dots,8\}$ are the gamma
matrices in the spinorial ${\bf 8_s}$ representation, and putting $B_{ij}$ in block-diagonal form by means of a
$SO(8)$ transformation, we have
\be
B_{ij}={\rm diag}\left[
\left(\begin{array}{cc}
   0 & J_1 \\
-J_1 & 0
\end{array}\right),\dots,\ \left(\begin{array}{cc}
   0 & J_4 \\
-J_4 & 0
\end{array}\right)
\right]\ ,
\ee
and similarly for $\hat B_{ab}$, with $\hat b_1,\dots, \hat b_4$ instead of $J_i$.
The non-vanishing elements of $\hat B_{ab}$ are related to the $J_i$'s by
\bea
&&\hat b_1={1\over 2}(-J_1+J_2+J_3+J_4)\ ,\qquad \hat b_2={1\over 2}(+J_1-J_2+J_3+J_4)\ ,
\nonumber\\
&&\hat b_3={1\over 2}(+J_1+J_2-J_3+J_4)\ ,\qquad \hat b_4={1\over 2}(+J_1+J_2+J_3-J_4)\ .
\label{hatba}
\eea
The anticommutation relations then become
\begin{equation}
\big\{ Q_a ,\ Q_b \big\} = \mathbb{I}\, \delta_{ab} \; P^0 +  \tilde \gamma^0\; \hat B_{ab} \ .
\end{equation}
Since $(\tilde \gamma^0)^2 = - 1$, the eigenvalues of $\tilde \gamma^0 \hat B$ are $\pm  \hat b_i$.
Therefore, the eigenvalues of the anticommutator matrix are $P^0 \pm \hat b_i$, $i=1,\ldots,4$.
In any unitary representation the matrix $\big\{ Q_a ,\ Q_b \big\}$ is definite positive, thus the BPS bound is
\begin{equation}\label{bound-jamesbound}
P^0 \ge \hat b_\text{max} \ .
\end{equation}
where $\hat b_\text{max}$ is the maximum of $\pm \hat b_i$.
%
%
%

$P^0$ generates translations in the time $t$. For the membranes considered in this paper lying at $r=0$, the proper time
is given by  $d\tau =\frac R 2 dt$, see (\ref{metric}). Therefore their energies $E$ are related to $P^0$ by  $E=2P^0/R$.
Defining $\eta_i =\sgn(J_i)$, the signs are subject to the condition $\eta_1\eta_2\eta_3=-\eta_4$.
This implies that, for these membranes, $\hat b_\text{max}$ is nothing but ${1\over 2}\sum_{i=1}^4 |J_i |$. Thus
the  energies of our membrane solutions are subject to the  bound
\begin{equation}\label{jamesbond}
E \ge \frac 1 R \sum_{i=1}^4 |J_i | \ .
\end{equation}
When the bound is saturated, the matrix of anticommutators have some zero eigenvalues, implying that some fraction of supersymmetry is preserved.

When three or more $J_i$ are non-vanishing and generic\footnote{When some $J_i$ have coincident values, some $\hat b_i$ will be equal to each other, implying the possibility of enhancement of supersymmetry.
However, it is easy to see that this possibility is not realized our membrane solutions subject to the condition $\eta_1\eta_2\eta_3=-\eta_4$.},
there is only one vanishing eigenvalue and the corresponding state saturating the bound preserves
1/8 of the supersymmetries. When two $J_i$ are non-zero and generic, there are two vanishing eigenvalues and the corresponding state preserves 1/4 of the supersymmetries.
Finally, states with only one non-zero $J_i$ have four vanishing eigenvalues and the solution preserves 1/2 of the supersymmetries.

\section{Energy and angular momenta }\label{sec:EJ}

\subsection{General formulas}

According to Noether's theorem, if
\be
^\epsilon X^\mu = X^\mu + \epsilon\; \delta X^\mu + o(\epsilon^2)\ .
\ee
is a continuum transformation with parameter $\epsilon$ such that $S[^\epsilon X, h] = S[X, h]$, then
\be
J^\alpha = \delta X^\mu\,\frac{\partial\cal L}{\partial\partial_\alpha X^\mu}\bigg|_\text{on-shell},
\ee
is a conserved current,
\begin{equation}
 \nabla_\alpha J^\alpha= 0\ ,
\end{equation}
and, therefore,
\be
Q \equiv \int\!\dd\sigma^1\dd\sigma^2\; J^0 =
-T_2\;\int\!\dd\sigma^1 \dd\sigma^2\;\sqrt{-h}\; h^{0\alpha}\; G_{\mu\nu}(X)\; \delta X^\mu\;\partial_\alpha X^\nu
\ ,
\ee
is a conserved quantity
\begin{equation}
 \frac{\dd Q}{\dd \sigma^0} = 0\ .
\end{equation}

If we apply this standard procedure to isometries of the background
(and hence symmetry transformations) we can define the following conserved charges.

\begin{description}
\item{Energy:}  $\;\;\frac{R}{2}\,\delta X^0 = \frac{R}{2}\,\delta t = -1\;$.
\be\label{energy}
E = V_2\;T_2\;R\;\frac{\omega_0}{2}\;\frac{h_c^{00}}{\sqrt{-h}}\; \ , \qquad
\;V_2\equiv\int\!\dd\sigma^1\dd\sigma^2 =4\pi^2\ .
\ee

\item{Angular momenta:}  $\;\;\delta \xi^i = 1\;, \forall i$.
\be\label{momentos}
J_i = V_2\;T_2\;\;R^2\; \frac{h_c^{0\alpha}}{\sqrt{-h}}\; \frac{\beta^i_\alpha}{2}\;\mu_i{}^2
= E\;\frac{R}{\omega_0}\;\mu_i{}^2\;\left(
\beta^i_0 + \frac{h_c^{01}}{h_c^{00}} \;\beta^i_1  +\frac{h_c^{02}}{h_c^{00}} \;  \beta^i_2\right)\ .
\ee
\end{description}

\subsection{Energy and momenta of non-collapsed membranes}

Evaluating these formulas on our family of solutions~(\ref{sngral})
\begin{subequations}
\begin{align}
E & = V_2\;T_2\; R^2\;\mu_1\;\mu_3\;\left(
\frac{2\,z - \mu_1{}^2 -\mu_2{}^2}{z-\mu_2{}^2}\;
\frac{2\,z - \mu_3{}^2 -\mu_4{}^2}{z-\mu_4{}^2}\;\frac{3\,z-1}{z}\right)^{\!1/2}\ ,\\
\eta_1\, J_1 &= R\;E\;\mu_1\; \left(\frac{z-\mu_1{}^2}{(3\,z-1)(z-z_0)}\;
\frac{z^2 - z_2\,z + (\mu_1{}^2 +\mu_2{}^2)\,z_0}{2\,z - \mu_1{}^2 -\mu_2{}^2}\right)^{\!1/2}\ , \\
\eta_2\, J_2 &= R\;E\;\mu_2\; \left(\frac{z-\mu_2{}^2}{(3\,z-1)(z-z_0)}\;
\frac{z^2 - z_2\,z + (\mu_1{}^2 +\mu_2{}^2)\,z_0}{2\,z - \mu_1{}^2 -\mu_2{}^2}\right)^{\!1/2}\  , \\
\eta_3\, J_3 &= R\;E\;\mu_3\; \left(\frac{z-\mu_3{}^2}{(3\,z-1)(z-z_0)}\;
\frac{z^2 - z_4\,z + (\mu_3{}^2 +\mu_4{}^2)\,z_0}{2\,z - \mu_3{}^2 -\mu_4{}^2}\right)^{\!1/2}\ , \\
\eta_4\, J_4 &= R\;E\;\mu_4\; \left(\frac{z-\mu_4{}^2}{(3\,z-1)(z-z_0)}\;
\frac{z^2 - z_4\,z + (\mu_3{}^2 +\mu_4{}^2)\,z_0}{2\,z - \mu_3{}^2 -\mu_4{}^2}\right)^{\!1/2}\ .
\end{align}
\end{subequations}
%
where we have introduced the signs of $\omega_2, \omega_4,m,n$ in the following way,
\be
\sgn(\omega_2) \equiv \eta_2\ , \qquad
\sgn(\omega_4) \equiv \eta_4 \ , \qquad
\sgn(m) \equiv -\eta_2\; \eta_1\ , \qquad
\sgn(n) \equiv -\eta_4\;\eta_3\ ,
\ee
so that $\sgn(J_i) = \eta_i$.

In the limit $z\to 0$ both $E$ and $J_i$ tend to infinity. It is straightforward to show that in this limit,
\be
J_i = \eta_i\, R\, E\,\mu_i{}^2\qquad\Longrightarrow\qquad E = \frac{1}{R}\;\sum_{i=1}^4 |J_i|\ .
\label{EJ}
\ee
This simple relation is due to the fact that in this limit the solution becomes supersymmetric, as explained in section 4 and  will be seen more explicitly in  section 6.

On the other hand, in the opposite limit  $|z|=\infty$, one finds the  solution with
$J_i = \eta_i\, \frac{R}{\sqrt{6}}\, E\,\mu_i$, giving
\be
E^2 = \frac{6}{R^2}\;\sum_{i=1}^4 J_i{}^2\ .
\ee
There is no preserved supersymmetry for this solution. The general relation between $E$ and $J_i$ for solutions with arbitrary $z$ is given in appendix A, for completeness.

\subsection{Energy and momenta of collapsed membranes}

Let us now consider the collapsed membrane configurations with $\vec m=K\, \vec n$.
In this case, the expressions (\ref{energy}), (\ref{momentos})
for $E$ and $J^\alpha$ become ambiguous and need a proper regularization.
The same ambiguity occurs for the BMN string if one attempts to compute the energy and angular momentum using
the Nambu-Goto action.
In this case, the solution describing the BMN state is $X^0=\omega_0\,\tau $, $\phi=\omega_0\,\tau $,
where $\phi $ is an angle of the
$S^5 $ sphere.
The proper way to do the calculation is, as in~\cite{GKP}, to use the Polyakov action in the conformal gauge,
and then compute $E$ and $J$ (obtaining $E \propto J $).
For membranes, there is no possibility of a conformal gauge.
The closer analog is the  gauge $h_{01}=h_{02}=0$ and $h_{00}=- ( h_{11}\, h_{22} -h_{12}{}^2)$.
The formulas (\ref{energy}), (\ref{momentos}) for the energy and angular momentum
in this gauge reduce to
\be
E= V_2\, T_2\, {\frac R 2}\, \omega_0 \ ,\qquad J_i=V_2\, T_2\, R^2\,\mu_i^2\, \omega_i\ .
\ee
In addition, for the collapsed membrane with $\vec m=K\vec n$, $g= 0$ and the constraint $h_{00}=- g$ implies the relation
\be\label{r2}
\omega_0 =2\;\sqrt{\sum_{i=1}^4\,\mu_i^2\;\omega_i^2}\ .
\ee
Since $\omega_0\neq 0$, at least one $\mu_i$ and $\omega_i$ must be non-vanishing. Taking
$\mu_1\neq 0$, $\omega_1\neq 0$, the equations of motion then imply the additional relation
\be\label{r1}
\omega_i{}^2=\omega_1{}^2=-\Lambda\ ,\qquad  \forall i\ {\rm such\ that}\  \mu_i\neq 0\ .
\ee
It follows that $\omega_0=2\;|\omega_1| $. This agrees with the general formulas of \cite{russo} particularized to the case $\vec m=K\vec n$.
Therefore
\be\label{EJcol1}
E= {\frac J R}\ ,\qquad J=\sum_i |J_i|\ .
\ee
In addition, the constraint associated with the gauge choice $h_{01}=h_{02}=0$ imposes the condition
\be\label{consgauge}
\sum_{i=1}^4 \, m_i\; J_i =0\ .
\ee

The derivation of the previous formulas implies dealing with membranes with  null world-volume, i.e. $h=0$, for which classical methods
are not, in general, justified. Indeed, these membranes can be more properly  viewed as the limit of large angular momentum of
 general  non-collapsed, regular membranes of the form (\ref{ansatz}). This is obviously the case as can be explicitly seen from the general formulas given in sect. 4 of \cite{russo},
 where the large $J$ limit indeed leads to the conditions (\ref{r2}), (\ref{r1}) and (\ref{EJcol1}) (while the condition (\ref{consgauge}) holds for any finite $J$).
 In general, one finds \cite{russo} $E=J/R+ O(1/J)$.

\section{Supersymmetry conditions for the solutions}

\subsection{Supersymmetry equations}

We shall now investigate the configurations of the form~(\ref{ansatz}) which preserve some fraction of supersymmetry.
A configuration preserves a supersymmetry for every independent Killing spinor  $\epsilon$ defined in (\ref{susan}) that
satisfies
\begin{equation}
\Gamma_\kappa\; \epsilon = \pm \epsilon \ , \qquad \Gamma_\kappa = \frac{1}{3! \sqrt{-h}} \epsilon^{a b c} \partial_a X^\mu \partial_b X^\nu \partial_c X^\rho \Gamma_{\mu\nu\rho}\ ,
\label{affr}
\end{equation}
where $\Gamma_\kappa$ is the $\kappa$-symmetry matrix, the Gamma matrices are given by~(\ref{gammas}),
and $+1 (-1)$ stands for the M2 (anti) brane.
Substituting the ansatz~(\ref{ansatz}) we find
%
%
\begin{equation}\label{gammakappa}
\Gamma_\kappa
= \frac{R^3}{8 \sqrt{-h}} \mu_i\, \mu_j\, \beta_1^{\,i} \beta_2^{\,j}    \Big(  \omega_0 \, \gamma_0
+ \mu_k \beta_k^{\,0} \gamma_{k+6} \Big) \gamma_{i+6}\, \gamma_{j+6} \ .
\end{equation}
where summation over $i$, $j$ is understood and $k$ indexes between 1 and 4.

%
Using~(\ref{susan}), we find that the Killing spinors of $\AdS_4 \times S^7$ must satisfy
\begin{equation}\label{susycond}
\mathcal M^{-1}\; \Gamma_\kappa\; \mathcal M\; \epsilon_0 = \pm\;\epsilon_0 \ .
\end{equation}
%

After some algebra, the full supersymmetry equations reduce to%
\footnote{This algebra requires commuting ${\cal M}$ with $\Gamma$ matrices.
Useful relations can be found in the appendix B of~\cite{japos}.}
\begin{equation}\label{fullbps}
\begin{aligned}
&\left\{ \sum_{i<j} \gamma_0 \gamma_{i+6} \gamma_{j+6} \Big( \omega_0 \mu_i \mu_j \beta_{12}^{i j} - \mu_i \mu_j \mu_k^2 \beta^{ij k} O_k \Big) M_t^2 M_i^2 M_j^2
\right. \\&\left.
- \gamma_{7,8,9,10}  \sum_{i j k l} \varepsilon_{i j k l} \beta^{ij k} \mathbb X_l
 \prod_{k=1}^4 \mu_k M_k^2  \right\} \epsilon_0
\\&\qquad
=
- \sqrt{ \omega_0^2 \sum_{i<j} \mu_i^2 \mu_j^2 (\beta_{12}^{i j})^2 - \sum_{i<j<k} \mu_i^2 \mu_j^2 \mu_k^2 ( \beta^{ij k} )^2 } \ \epsilon_0\ ,
\end{aligned}
\end{equation}
where
\begin{equation}
\begin{aligned}
O_k & = \hat\gamma\; \gamma_0\; \mathbb X_k \ , \\
O_4 & = - O_1\; O_2\; O_3 \ ,
\label{sonn}
\end{aligned}
\end{equation}
the $ \mathbb X_k$ have been defined in (\ref{matX}), and
\begin{equation}
\begin{aligned}
\beta_{\alpha\beta}^{i j} & \equiv \beta_\alpha^{i} \beta_\beta^{j} - \beta_\alpha^{j} \beta_\beta^{i}\ , \\
\beta^{ijk} & \equiv \beta_0^{i}\; \beta_{12}^{jk} + \beta_0^{j}\; \beta_{12}^{ki} + \beta_0^{k} \beta_{12}^{ij}\ .
\end{aligned}
\end{equation}

Equation~(\ref{fullbps}) is highly complicated in general.
However, for our ansatz~(\ref{ansatz}), (\ref{abo}), gets simplified in a striking way.
%
%
In particular, it is easy to check that  all the $\beta_{12}^{\ i j}$ either vanish or are proportional to
\begin{equation}\label{eqN}
N = a d - b c \ .
\end{equation}
This implies that both terms of~(\ref{fullbps}) are proportional to $N$.
On the face of it, it might seem that if $N=0$ then the supersymmetry condition~(\ref{fullbps})
is trivially satisfied
for all $32$ spinors  $\epsilon_0$. However, the $N=0$ case is rather subtle, because in this case
$\vec m$ is proportional to $\vec n$
and the M2 brane collapses to a string-like configuration, as explained at the end of section 3.2.
In this case the equation~(\ref{affr}) becomes singular and cannot be used. We will return to this case in section 6.3.

\subsection{Supersymmetry of the non-collapsed membranes}\label{8thsusysol}

We first investigate  the supersymmetry conditions for $N\neq 0$, for generic values of the $\mu_i$'s.
Let  $\eta_k$ denote the eigenvalues of the $O_k$ operators,
\begin{equation}\label{etas}
 O_k\; \epsilon_0 = \eta_k\; \epsilon_0\ \ ,\qquad  k=1,2,3\; .
\end{equation}
Since $O_k^2 = 1$, the eigenvalues are just equal to $\pm 1$.
This leads to only three independent conditions, since $\eta_4 = - \eta_1 \eta_2 \eta_3$ (see equation (\ref{sonn})).
With no loss of generality we can set $\eta_1=\eta_2=\eta_3=1$, $\eta_4=-1$, since the sign of any $\eta_i$  can be
reversed by  a coordinate redefinition $\xi^i\to -\xi^i $.
Let us start by fixing,
\begin{align}
\alpha & = - \frac{\mu_1}{\mu_2}
\ , &
\beta & = \frac{\mu_3}{\mu_4}
\ ,
\end{align}
By using~(\ref{etas}), the supersymmetry condition~(\ref{fullbps})
leads to two equations
\begin{subequations}\label{C1C2}
\begin{align}
\mu_1^2 \;\omega_{1} + \mu_2^2\;  \omega_{2} & = \frac12\ \,  \omega_0\; ( \mu_1^2 + \mu_2^2 ) \ , \\
\mu_3^2\;  \omega_{3} -\mu_4^2\; \omega_{4} & = \frac12\ \, \omega_0\; ( \mu_3^2 + \mu_4^2 ) \ .
\end{align}
\end{subequations}
Note that these equations only restrict the possible values of the parameters, but they do not imply any condition on the spinor.
Therefore  equations (\ref{C1C2}) do not reduce the number of supersymmetries.
Once (\ref{C1C2}) are imposed on the parameters, both sides of the supersymmetry equation~(\ref{fullbps})  become identically  zero.
In terms of the coordinates $\tilde \sigma^\alpha $, the solution takes the simple form
\begin{equation}\label{cans}
\begin{aligned}
t & = \omega_0\; \tilde \sigma^0\ ,\qquad & r & = 0 \ ,
\qquad \mu_i  = \text{constant}  , \\
 \xi^1 & =\tilde \sigma^1 \ , &
 \xi^2 & =  \tilde \omega_2\; \tilde \sigma^0 +\tilde m\; \tilde\sigma^1 \ , \\
 \xi^3 & = \tilde \sigma^2 \ , &
 \xi^4 & = \tilde \omega_4\; \tilde \sigma^0 + \tilde n\; \tilde \sigma^2 \ .
\end{aligned}
\end{equation}
with
\be\label{tdtd}
\tilde \omega_{2}  =\frac12\  \omega_0 \; \left( 1+ \frac{\mu_1^2}{ \mu_2^2 }\right) \ , \qquad
\tilde  \omega_{4}  =- \frac12\ \omega_0 \; \left( 1+\frac{\mu_3^2 }{ \mu_4^2 }\right)\ ,
\qquad \tilde m = -\frac{\mu_1^2}{\mu_2^2}\ , \qquad \tilde n =+\frac{\mu_3^2}{\mu_4^2}\ .
\ee
On shell (i.e. upon use of (\ref{sngral})), this M2 brane has a singular induced metric, $h=0$.
Nonetheless, it should be noted that the membrane is regular, in particular, it is not collapsed to a string,
despite the fact that the induced world-volume metric has vanishing determinant $h=0$.
The  phenomenon is similar to the one  found for strings in~\cite{Mateos}.
The interpretation is that these configurations describe tensionless membranes, since the world-volume is null. Physically, it means that, for these membranes,
the energy due to the tension is negligible compared to the energy due to rotation (see also section 5).

In conclusion, the M2 brane configuration (\ref{cans}) is supersymmetric for  Killing spinors satisfying the three conditions~(\ref{etas}). Therefore
our solution preserves 1/8 of the supersymmetries of the background.
Furthermore, we note that the solution  is just the $z=0$ solution to the equations of motion given in $(\ref{snz=0})$.

The number of supersymmetries can also be deduced from the BPS algebra.
For generic values of $\mu_i$'s, the bound (\ref{jamesbond}) is saturated with the four $J_i$ non zero and
different from each other, as shown in section 5.2, see (\ref{EJ}).
In this case the $8\times 8$ matrix $\{ Q_a,Q_b\} $ has a unique zero eigenvalue, hence only 1/8 of the
supersymmmetries is preserved, in agreement with the above counting using the $\Gamma_\kappa$ matrix.

\subsection{Supersymmetry of the collapsed membranes}\label{susycolmem}

In terms of the new world-volume coordinate $\sigma\equiv \sigma^2+K\,\sigma^1 $, the solution for the M2 brane
collapsed to a string is obtained from the  ansatz~(\ref{ansatz}) by simply setting
$n_i=0$. This gives
\begin{equation}\label{ansatz-collapsed}
\begin{aligned}
t & = \omega_0\, \tau \ , &  r & = 0 \ , \\
 \mu_i & =  \text{constant} \ , &
\xi^i & = \omega_i\; \tau + m_i\; \sigma \equiv \frac12\; \beta_a^{i} \, \sigma^a\ ,
\end{aligned}
\end{equation}
where $\sigma^ a = (\sigma^0,\sigma^1)\equiv (\tau,\,\sigma)$, $a=0,1$.

The simplest way to study the supersymmetry of the collapsed membrane configuration is from the supersymmetry algebra.
In section 5.3 we have seen that these configurations saturate the BPS bound (\ref{jamesbond}),
and therefore they are all supersymmetric.
The preserved fraction of supersymmetries depends on how many $J_i$ are different from zero:

\medskip

\begin{itemize}

\item In the case of rotation in four planes with generic $J_i$'s non-zero, the
$8\times 8$ matrix $\{ Q_a,Q_b\} $ has  only one  zero eigenvalue.
As a result, the solution preserves 1/8 of the supersymmetries.

\item In the  case of rotation in three planes, only one of the $J_i$ vanishes, say $J_4$. Generically, the $\hat b_i$ given in (\ref{hatba}) are still   different from each other  and
 as a result the  matrix $\{ Q_a,Q_b\} $ has still  only one  zero eigenvalue. This solution also preserves 1/8 of the supersymmetries.

 \item In the  case of rotation in two planes, two of the $J_i$ vanish, say $J_3,\ J_4$. From (\ref{hatba}) we obtain $\hat b_1=-\hat b_2$ and
$\hat b_3=\hat b_4$.
It is easy to see that in this case
 the  matrix $\{ Q_a,Q_b\} $ has   two zero eigenvalues, coming from $E\pm {2\over R}\hat b_{3,4}  $ or $E\pm {2\over R} \hat b_1$, $E\mp {2\over R}\ \hat b_2$,
according to the signs of $J_1, J_2$. This solution  preserves 1/4 of the supersymmetries.

\item Finally, in the case of rotation in one plane, taking e.g. $J_2=J_3=J_4=0$,
there are four vanishing eigenvalues when the BPS bound is saturated. The membrane preserves 1/2 of the supersymmetry.
However, in this case the constraint (\ref{consgauge}) $\sum_{i=1}^4 m_i J_i =0$ implies that $m_1=0$: the membrane collapses to a point.
This is a BMN state.


  \end{itemize}

The case of the M2 brane (\ref{ansatz-collapsed}) collapsed to a string-like configuration
the $\kappa $-symmetry matrix $\Gamma_\kappa $ of the M2 brane is singular and cannot be used to determine the unbroken
supersymmetries.
The same problem exists for  strings collapsing to a point, like in the BMN solution~\cite{BMN,GKP}, representing a
collapsed string moving around the equator of $S^5$ at the speed of light; the
$\Gamma_\kappa $ matrix of the string is singular but one can use the supersymmetry algebra in a similar way as we did above
to show that the solution preserves 1/2 of the supersymmetries (see e.g.~\cite{Mateos}).
%
%

In the present case, since the membrane is collapsed to a string, one may try to determine the unbroken supersymmetries
by using the $\Gamma_\kappa $  matrix corresponding to an effective string.
In  appendix B we show that this approach reproduces the correct number of supersymmetries
obtained above from the  supersymmetry algebra.

\section{Generalization to $\AdS_4 \times S^7/\mathbb{Z}_k$}

The supersymmetric M2 brane configurations described in the previous sections admit a straightforward
generalization to the case of $\AdS_4 \times S^7/\mathbb{Z}_k$.
As explained above, the $\mathbb{Z}_k$ orbifold acts on the $\xi_i$ angles by   identification $\xi_i \sim \xi_i + {2\pi}/{k}$.
The spectrum on $\AdS_4 \times S^7/\mathbb{Z}_k$ is obtained by the projection of the original
spectrum on $\mathbb{Z}_k$ invariant states.
This leads to the following quantization conditions on the winding numbers:
\begin{equation}
m_i ,\ n_i \in \mathbb{Z} / k \ .
\end{equation}
Dimensional reduction of $\AdS_4 \times S^7/\mathbb{Z}_k$ along the $y$ coordinate gives the $\AdS_4 \times CP^ 3$ space (see section 2).
Finding novel supersymmetric states in this space is of particular interest given the connection with  ABJM theory.
To proceed, we recall that $y$ is the diagonal part of the four $\xi_i$ angles,
\begin{equation}
y = \frac14\big( \xi_1 + \xi_2 + \xi_3 + \xi_4 \big) \ .
\end{equation}
For our general ansatz~(\ref{ansatz}) this gives
\begin{equation}\label{ygeneral}
y= \omega_y \,\sigma^0 + m_y \,\sigma^1 + n_y\, \sigma^ 2\ ,
\end{equation}
where we have defined
\begin{align}
\omega_y & = \frac14 \sum_i\omega_i\ , &
m_y & = \frac14 \sum_i m_i\ , &
n_y & = \frac14  \sum_i n_i\ .
\end{align}
The other coordinates $\psi, \varphi_1,\ \varphi_2$ in eq. (\ref{artan}) have a similar $\sigma^0,\ \sigma^1,\ \sigma^2 $ dependence.
As pointed out in ~\cite{russo}, these type of configurations in the generic case
correspond to non-perturbative objects in the type II string theory.
Generally, in ten dimensions these configurations represent  bound states of D0 branes, D2 branes  and rotating circular fundamental strings.
The D0 brane charge arises from the momentum in the $y$ direction, $P_y=k\,\omega_y $.
Because the circles $\psi,\varphi_1,\varphi_2$ are contractible,
the net D2 brane and fundamental string charges are zero (just like the fundamental strings of~\cite {ART}).

Consider in particular the 1/8 supersymmetric non-collapsed M2 brane solution (\ref{cans}). In this case
\begin{equation}\label{yparticular}
4\,y= (\omega_2+\omega_4)\, \sigma^0 + (1+m)\, \sigma^1 + (1+n)\, \sigma^ 2\ ,
\end{equation}
where we removed tildes. Using (\ref{tdtd}), we see that D0 brane charge
$P_y= \omega_0\, k\, (\mu_1^2\mu_4^2-\mu_3^2\mu_2^2)/(8\mu_2^2\mu_4^2)$ is determined in terms
of angles $\mu_i$ representing the location of  the bound state system.

Consider now the collapsed membrane configurations of section 6.3. They are of the form
\begin{equation}
4\,y_\text{coll.} = \omega_y\, \sigma^0 +  n_y\, {\sigma^2}\ ,
\end{equation}
where we now use $\sigma^2$, instead of $\sigma$, to avoid possible confusion with the world-sheet string coordinate $\sigma $ of type IIA string theory.
The other coordinates  $\psi, \varphi_1,\ \varphi_2$ depend only on $\sigma^0$ and $\sigma^2$ as well. The configuration has non-vanishing D0 brane charge.
To see this explicitly, we recall that
another consequence of the orbifold projection is that the momentum along the $y$ is quantized as
$J_1 + J_2 + J_3 + J_4 = k p$,
%
for $p$ units of D0 brane charge (see related  discussion in ~\cite{japos}).
There are two cases to be distinguished:
\smallskip

\noindent a) $n_y=0$. In this case the string-shaped membrane is not wrapped around the eleven dimensional circle $y$. The other coordinates
$\psi,\varphi_1,\varphi_2$ will generically depend on $\sigma_2$, which, in this particular case, can  be identified with the string world-sheet coordinate.
The configuration then represents a bound state system of $p$ D0 branes and fundamental strings with vanishing total charge.

\smallskip

\noindent b) $n_y\neq 0$. In this case the string-shaped membrane is now wrapped around the eleven dimensional circle $y$.
As a result, upon reduction, the configuration does not contain any fundamental string, but it has $p$ units of D0 brane charge.

\medskip

It would be interesting to identify the dual BPS operators of ABJM three dimensional $\mathcal N = 6$ Chern-Simons theory, both for the collapsed membranes and for the 1/8 supersymmetric
M2 brane (\ref{cans}). In general, these operators
have conformal dimension $ kp/2$ and (like the configurations of~\cite{berenstein})  are  to be given in terms
of configurations  involving  non-abelian degrees of freedom in some non-trivial way.

\section{Giant Diabolo}

\subsection{BPS equation}

In this section we study a different class of supersymmetric membranes that also extend to the $\AdS_4$ part of the background.
It is convenient to introduce cylindrical coordinates.
The metric and three-form become
\begin{equation}\label{fullmetric_cil}
 \begin{aligned}
  \dd s^2 & = \frac{R^2}{4} \left\{ - \big( 1 + z^2 + \rho^2 \big)\ \dd t^2  + \dfrac{ (z\, \dd z + \rho\, \dd \rho)^2}{(z^2+\rho^2)(1+z^2+\rho^2)} + \dfrac{ (z\, \dd \rho - \rho \, \dd z)^2}{z^2+\rho^2} +\rho^2 \dd\varphi^2 \right\}
\\ & \quad +
R^2 \left\{ \dd\alpha^2 + \cos^2\alpha\; \dd\beta^2 + \cos^2\alpha\; \cos^2\beta\; \dd\gamma^2 +
\sum_{i=1}^4\, \mu_i^2\; \dd\xi_i^2 \right\}\\
C^{(3)} & = \frac{R^3}{8}\;\rho\;dt\wedge (z\,d\rho - \rho\,dz)\wedge d\varphi .
\end{aligned}
\end{equation}
The ansatz is as follows
\begin{equation}\label{ansatzdiabolo}
\begin{aligned}
t & = \sigma^0 \quad , \quad\qquad z = \sigma^2 \ ,\\
\varphi & = \alpha_0\; \sigma^0 + \alpha_1\; \sigma^1 + \alpha_2\; \sigma^2\ , \\
\xi^i & = m_i\;\left(s_0\;  \sigma^0 + s_1\;  \sigma^1 +s_2\;  \sigma^2\right)\ ,\\
\rho & = \rho(\sigma^2) \ .
\end{aligned}
\end{equation}
We shall derive the BPS equations using again the condition (\ref{affr}) on the background spinors.
We first  decompose the $\kappa$-symmetry matrix in two factors
\begin{align}
\Gamma & = - \tilde\gamma\; \tilde\Gamma\ ,  &
\tilde\gamma &\equiv \dfrac{f'\, \gamma_1 + (z \rho' - \rho)\, \gamma_2}{r \,(1 + {\rho'}^2 - {f'}^2)^{1/2}} \ , &
\tilde\Gamma &\equiv \dfrac1{\sqrt {-h}} \left( \sum_{i=1}^4 \delta_i + \tilde\delta \right) \ ,
\end{align}
where $f^2(z)\equiv 1 + \rho^2(z) + z^2 $, and
\begin{subequations}
\begin{align}
\delta_1 & = \frac12 \,\alpha_1\, \rho \,f\, \gamma_{03}\ , \ \qquad
\delta_2  = s_1\, f\, \gamma_0\, \sum_{i=1}^4\, \mu_i\, m_i\, \gamma_{i+6} \ , \qquad
\delta_3  = (\alpha_0\, s_1 - \alpha_1\, s_0) \,\rho\, \gamma_3\, \sum_{i=1}^4\, \mu_i\, m_i\,\gamma_{i+6}\ , \\
\tilde \delta & = (\alpha_1\, s_2 - \alpha_2\, s_1)\, \frac{\rho\, f}{(1+{\rho'}^2 - {f'}^2)^{1/2}}\;
\gamma_{30}\; \tilde\gamma\; \sum_{i=1}^4\, \mu_i\, m_i\,\gamma_{i+6} \ .
\end{align}
\end{subequations}

An important feature of these matrices is that $\tilde\gamma$ and $\tilde\Gamma$ do not commute unless $\tilde\delta = 0$.
We will assume $\tilde\delta = 0$ in order to get an analytic solution. Thus we take
\be
 s_i = a\; \alpha_i\ ,\qquad i=1,2\label{conds}\ ,
\ee
with $a$ arbitrary. Once this condition is implemented, our ansatz becomes equivalent to the ansatz considered in~\cite{japos} using spherical coordinates.
The cylindrical coordinates are more convenient to exhibit how
the various geometries are realized for different values of the parameters.
Some of the geometries shown here are novel.

The supersymmetry condition is
\begin{equation}
\left[ \mathcal{M}^{-1} \tilde\gamma \mathcal{M} \ \mathcal{M}^{-1} \tilde\Gamma \mathcal{M} + \epsilon \right] \epsilon_0 = 0 \ .
\end{equation}
with $\epsilon = +1 (-1)$ for the (anti) $M2$ brane.
In order to cancel out terms proportional to $M_i^2 M_j^2$, for $i\neq j = 1, 2, 3, 4$, we demand
\bea\label{consdiab1}
m_i &=& e_i\; m\quad,\quad e_i^2=1 \quad ,\qquad m>0\quad,\quad \qquad i=1,2,3, 4\ ,\cr
\mathbb X_i\; \mathbb X_4\;\epsilon_0 & =& - e_i\;e_4\;\epsilon_0 \qquad,\qquad\qquad i=1,2,3 \ .
\eea
Note that we can impose these conditions on the spinor since $\mathbb X_i^2 = -1$ and $[\mathbb X_i,\ \mathbb X_j] = 0$.
However, due to the relation $\prod_{i=1}^3 (\mathbb X_i\, \mathbb X_4)= +1$ (see \ref{matX}), we have 
that $\prod_{i=1}^4\,e_i = -1$.
Furthermore, this also implies that there are just two independent constraints in (\ref{consdiab1}).
The signs $e_i $ can be reabsorbed into a redefinition of $\xi^i$, and $m$ can be absorbed by the $s_\alpha$'s.
Therefore in what follows we set $m_1 =m_2=m_3=-m_4=+1$.
We omit some details of the computation, which is straightforward, albeit tedious.
The resulting conditions turn out to be
\begin{align}
\gamma_{0,10}\; \epsilon_0 & = \eta_1\; \epsilon_0 \qquad , \qquad \gamma_1\; \epsilon_0 = \eta_2\; \epsilon_0 \ ,
\label{consdiab2} \\
 s_0 & = a \; \alpha_0 +  a\;\eta_2\; \left( 1 - \frac1b \right) \ , \\
 \epsilon\ \sgn \{ \alpha_1 \} & = \sgn \left\{ \dfrac{ (\rho f)^2 + b^2  f^2 - \rho^2 \left( b - 1 \right)^2 }{ \rho^2 + b (1+z^2) }\right\}\ , \label{signcond} \\
 \rho' & = (b-1)\; \dfrac{z\; \rho(z)}{ \rho^2 + b \, (1+z^2)}\ ,\label{bpseq}
\end{align}
where $b \equiv  - 2\, \eta_1\, \eta_2\, a$.
Equation~(\ref{bpseq}) is the BPS  differential equation that determines the shape of the M2 brane.
Upon imposing these conditions, the solution takes the form
\begin{subequations}\label{sndiabolo}
\begin{align}
t & = \sigma^0 \ , \ \qquad
\varphi  = w \sigma^1  \ ,\qquad z=\sigma^2 \ ,\\
\xi^1 = \xi^2 = \xi^3 = -\xi^4 & =  \frac {\eta_1} {2}\, ( 1- b )\,  \sigma^0 -
\frac 1 2\, \eta_1\,\eta_2 \, b\, \alpha_1\, \sigma^1 \ .
\end{align}
\end{subequations}
where we have made a coordinate redefinition
$ \alpha_0 \sigma^0 + \alpha_1 \sigma^1 + \alpha_2 \sigma^2\to w \sigma^1$
($w$ is a winding number).
The general solution of equation~(\ref{bpseq}) is  given by
\begin{equation}\label{solucionjorge}
z^2 = r_0{}^2 \;\rho^c - 1 - \rho^2 \ ,
\end{equation}
where $r_0$ is an integration constant and
\begin{equation}
c = \frac{2\,b}{b-1} \ .
\end{equation}
The solution is symmetric  under $z\to -z$ and
it has an important feature: it always crosses the $z=0$ hyperplane smoothly. To see this, we differentiate the equation~(\ref{solucionjorge}) with respect to $\rho$,
\begin{equation}\label{difz}
  2 \,z \;\frac{\dd z}{\dd \rho} = c\; r_0{}^2\; \rho^{c-1} - 2\, \rho \ ,
 \end{equation}
and consider the limit $z\to 0$. From (\ref{solucionjorge}), it is easy to see that in this limit the r.h.s. of (\ref{difz}) does not vanish.
Therefore, when $z\to 0$, one has  $\partial_\rho z \to \infty$ (or $\rho'(0)=0$),
which is the required condition for a smooth transition.

{} From (\ref{consdiab1}) and (\ref{consdiab2}) it follows that the spinor must satisfy four independent, compatible
constraints, so the solution will preserve at least $\frac{1}{16}$ of the supersymmetries. 
The condition~(\ref{signcond}) is non-trivial, but it can be shown that the  solutions described below do satisfy it.

\subsection{Brane scanning}

\begin{figure}[t]
\centering
\includegraphics[width=15.5cm]{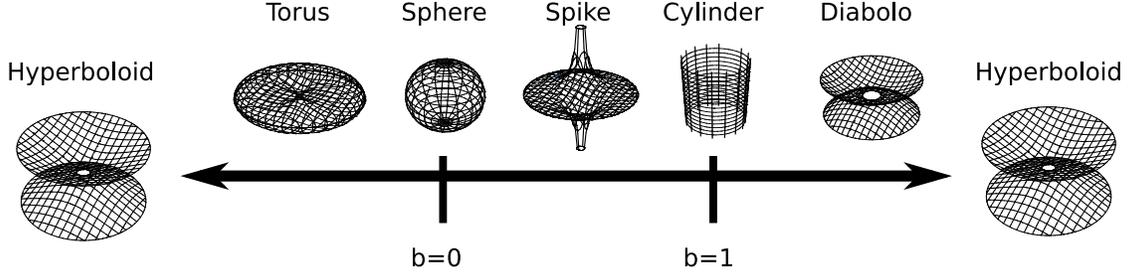}
\caption{The solution~(\ref{solucionjorge}) describes different geometries depending on the value of the parameter $b$.\label{shapes}}
\end{figure}

The solution~(\ref{solucionjorge}) describes membranes of diverse geometries depending on the value of the constant $b$, as shown in figure~\ref{shapes}. Generically, these membranes have angular momenta $J_\varphi,\ J_i$ in the $\varphi $ and $\xi^i $ directions (general formulas are given in appendix C).
 The standard expression for the Hamiltonian in the gauge $t=\sigma^0$ leads to the general formula
\be\label{japonE}
E={1\over R}\Big( 2|J_\varphi| +\sum_{i=1}^4 |J_i|\Big)+ \frac{2T_2}{R} \int d\sigma_1 d\sigma_2 \ {\cal L} \ .
\ee
As shown in \cite{japos}, for these  solutions the Lagrangian ${\cal L}$ becomes  a total derivative. This implies  that 
the last term vanishes in the case of M2 branes without boundaries. The resulting energy saturates the bound that one finds from the superalgebra, which
in case of rotation in both  $\AdS_4$ and $S^7$ is a slight generalization of the results of section 4.
Indeed, the term $ 2 |J_\varphi| $ comes from the contribution
$\tilde\gamma^0 \tilde\gamma_{23} M^{23} = \tilde\gamma^0 \tilde\gamma_{23} J_\varphi $.
Since the matrices $  \tilde\gamma^0 \tilde\gamma_{23}$ and   $ \tilde\gamma^0  $
commute, they can be simultaneously diagonalized and the eigenvalues of the matrix $\{ Q_a, Q_b \} $ are  $P^0 \pm J_\varphi \pm \hat b_i $, leading
to the bound $P^0\geq {1\over 2}\Big( 2|J_\varphi| +\sum_{i=1}^4 |J_i|\Big)$.

{}For uncompact M2 branes, the last term in (\ref{japonE})  will give a non-vanishing contribution to the energy.

\subsubsection{Giant spherical graviton}\label{sphere}

This appears for $b=0$ (which implies $c = 0$).  The solution (\ref{solucionjorge}) then becomes
\begin{equation}\label{solutionsphere}
 z^2+\rho^2 = r_0{}^2 - 1\ ,
\end{equation}
which is the equation of a sphere of radius $R_0\equiv \sqrt{r_0{}^2 - 1}$ in   cylindrical coordinates. In this case our ansatz reads
\be
t  = \sigma_0\quad ,\quad \varphi  =  w \, \sigma_1\quad, \quad z = \sigma_2\quad ,
\quad \xi^i = {\eta_1 \over 2} \ \sigma^0 \ .
\ee
{}From the formulas of appendix C one finds that this solution has $J_\varphi=0$ and $J_i = \eta_1\, e_i\, e_4\,\mu_i{}^2\,\pi\, |w| \,T_2\,R^3\,R_0$.
The energy is $E= \pi\, |w| \,T_2\,R^2\,R_0$, and therefore $E= \frac{1}{R}\;\sum_{i=1}^4\,|J_i|$.

\subsubsection{Cylinder}

{}For $b=1$, the constant $c$ tends to infinity and the solution~(\ref{solucionjorge}) is no longer valid. We have to return to the original equation~(\ref{bpseq}),
which now gives $\rho' = 0$, so the  radius $\rho=\rho_0$ of the cylinder is constant and arbitrary.
Using the formulas of appendix C
it can be easily shown that this is the only case where the angular momenta $J_\varphi,\ J_i$ vanish.

The energy is $E=\frac12\, T_2\,\pi\,|w| \, R^2\,L$, where $L$ regularizes the (infinite) length in $z$ direction.
Note that it is independent of $\rho_0$,
i.e. expanding the cylinder does not cost any energy, which is   consistent with the fact that solution exists for arbitrary radius.
This is  why the M2-brane can be in equilibrium in spite of the fact that $J_\varphi=J_i=0$.

\subsubsection{Giant spike}
This appears for $0 < b < 1$. In this interval for $b$, the constant $c$  covers all the negative real numbers, $c < 0$, so we can write it as $c = -|c|$, and the solution becomes
\begin{equation}
 z^2 = \dfrac{r_0^2}{\rho^{|c|}} - 1 - \rho^2 \ .
\end{equation}
At $z=0$, $\rho$ has a unique, non-vanishing value. As shown above, the transition between $z>0$ and $z<0$ is smooth.
At $z\to\pm\infty$, one has $\rho\to 0$, and this solution takes the form of a bulb with a spike,
which  in the dimensionally reduced theory can be interpreted as an open string stretched to infinity.
This solution was  found by Nishioka and Takayanagi in~\cite{japos}.  Now the energy picks a contribution
from the boundary at infinity: $E= \frac{1}{R}\Big( 2|J_\varphi| +\sum_{i=1}^4 |J_i|\Big)+ \frac12\, T_2\,\pi\,|w| \, R^2\,L$.

\subsubsection{Hyperboloid}\label{hyperboloid1}

In the limit that $b$ tends to $\pm $ infinity, the exponent $c$ in (\ref{solucionjorge})
approaches the fixed value  $c=2$. The solution (\ref{solucionjorge}) now reads
\begin{equation}
 (r_0^2-1) \rho^2 -z^2 = 1 \ ,
\end{equation}
which is the equation of an hyperboloid. Note that $r_0^2 >1$ for a real solution.
It has  finite $J_\varphi/(bL) ,\ J_i/L$ with $2|J_\varphi | /\sum_i|J_i| = |b| \to\infty $.

\subsubsection{Giant Diabolo}

\begin{figure}[tbh]
\centering
\subfigure[General view of the giant diabolo.]{\includegraphics[width=6.5cm]{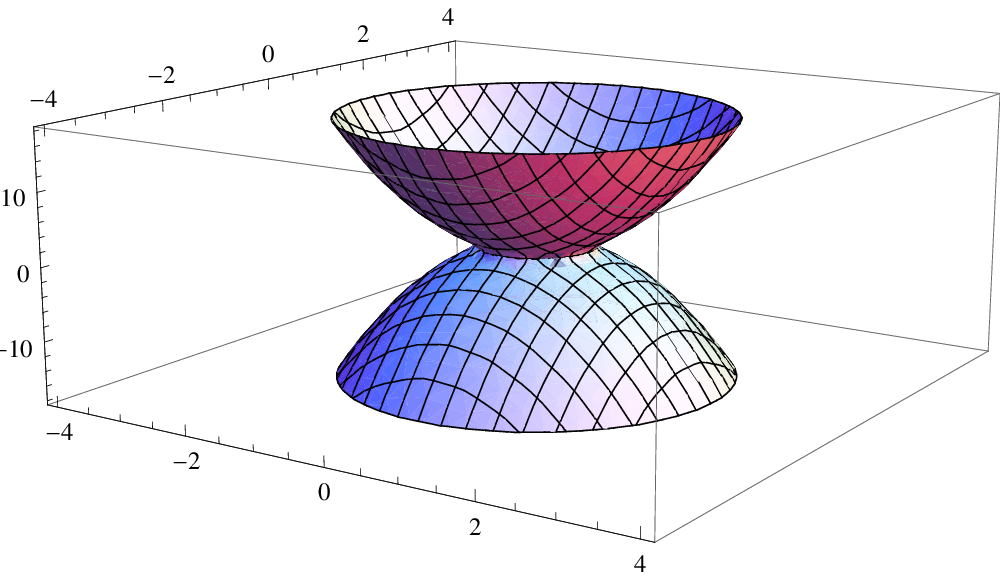}}
\subfigure[In this close up we can see that the transition between the two lobes is smooth.]{\includegraphics[width=6.5cm]{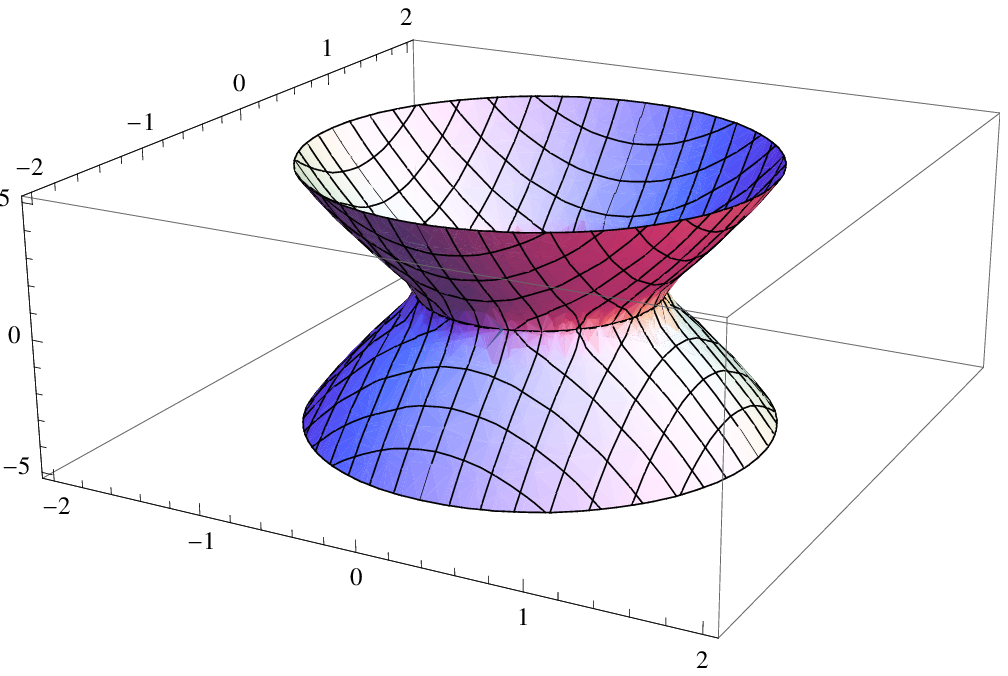}}
\caption{The giant diabolo for $b=1.8$ and $r_0^2 = 6$.\label{figdiabolo}}
\end{figure}

Consider now $b>1$. Then the exponent $c$ in (\ref{solucionjorge}) is always greater than two.
At $z=0$, $\rho$ again takes a unique, finite value. At $z\to\pm \infty$, the $\rho^c$ term  dominates and
\begin{equation}
 z \sim \pm r_0 \rho^{c/2}\ ,
\end{equation}
where $c/2>1$. This geometry resembles the shape of a diabolo, as can be seen in figure~\ref{figdiabolo}.
A difference with the hyperboloid ($c=2$) is that the diabolo exhibits a transition between negative and positive curvature at a certain value of $\rho$.
We recall that the solution has angular momentum both in $\varphi $ and $\xi^i $ directions and the general
formula for the energy is $E= \frac{1}{R}\Big( 2|J_\varphi| +\sum_{i=1}^4 |J_i|\Big)+ \frac12\, T_2\,\pi\,|w| \, R^2\,L$.

\subsubsection{Giant torus}

Now we assume a negative, but finite, value of the constant $b = - |b|$. In this case,
\begin{equation}
 c = \frac{2 |b|}{|b|+1} < 2 \ ,
\end{equation}
and solution (\ref{solucionjorge}) becomes
\begin{equation}
 z^2 = r_0^2 \rho^c - 1 - \rho^2 \ ,\qquad 0<c<2\ .
\label{torito}
\end{equation}
Since $c < 2$, the last term, which has a negative coefficient,  dominates at large $\rho$. Therefore $z^2$ can be positive only for $\rho $ less than some maximum value $\rho_M$,
where $z^2 = 0$. Similarly, the presence of the ``$-1$" on the r.h.s. shows that $\rho$ cannot be below a certain minimum value $\rho_m$, where $z=0$ again.
In short, when $0<c<2$, $z^2>0$ implies
that $\rho$  takes values in a finite interval  $[\rho_m,\ \rho_M]$. Since $z$ is a continuous function, it will have a maximum in this range.
In conclusion, eq. (\ref{torito}) represents a
torus-like geometry. This
is, indeed, the giant torus configuration found in~\cite{japos}. Being a compact M2 brane, one finds the simple relation  $E= \frac{1}{R}\Big( 2|J_\varphi| +\sum_{i=1}^4 |J_i|)\Big)$.

\bigskip

In conclusion, the solutions depends on two parameters, $r_0$, that characterizes a scale, and $b$. As the parameter $b$ is varied from $-\infty $ to $\infty$
one witnesses different transitions of the geometry, as illustrated by figure 1.

\section{Summary}

Summarizing, in the first part of this paper (sects. 3--7), we have investigated the following class of solutions
\be
t=\omega_0\sigma^0\ ,\ \ \ r=0\ , \qquad  Z_i=R \ \mu_i \ e^{i(\omega_i \sigma^0 + m_i \sigma^1 + n_i \sigma^2)}\ .
\ee
We identified two subclasses of supersymmetric solutions:
\medskip

\begin{enumerate}
\item Supersymmetric ``regular" M2 brane solutions
\be
 Z_1=R \ \mu_1 \ e^{i \sigma^1 }    , \ \ \  Z_3=R \ \mu_3 \ e^{i \sigma^2 }
 , \ \ \   Z_2=R \ \mu_2 \ e^{i (\omega_2 \sigma^0 +m\sigma^1) }
 , \ \ \   Z_4=R \ \mu_4 \ e^{i (\omega_4\sigma^0+n \sigma^2 )} ,
\ee
with $m,\ n,\ \omega_2,\ \omega_4$ determined in terms of $\mu_i$ (up to signs). They are tensionless and non-collapsed; for generic values of the parameters
they preserve 1/8 of the supersymmetries.

\medskip
\item  1/4 and 1/8 supersymmetric collapsed M2 brane solutions
\be
 Z_i=R \ \mu_i \ e^{i (\omega_i  \sigma^0 + m_i\sigma^1) }   \ ,
\ee
where the amount of preserved supersymmetries depends on the values of the parameters. The different cases
were analyzed in  detail in section 6.3 and in appendix B. The parameters are subject to the relations (\ref{r1}) and (\ref{consgauge}).
\end{enumerate}

\medskip
As discussed, the solutions admit globally non-trivial generalizations, which can be obtained by redefinitions of $\sigma^i$.

In all cases, supersymmetry is achieved in the same limit where $E, J\to\infty $ and the M2 branes become tensionless,
i.e. the determinant of the world-volume metric vanishes. This is the analog of the phenomenon found in~\cite{Mateos} for strings.
The main difference between the configurations 1 and 2 is that, in the first case,
the M2-brane extends in two directions, which wrap around $\xi^i $ coordinates, while in the second case the membrane is collapsed to a string and
extends in a single direction.

Our configurations have also some similarity with the BMN configurations~\cite{BMN} in the sense that in both cases they correspond to  null objects
moving around circles of $S^5$ or $S^7$, with $E\propto J\to\infty$.
It would be interesting to see if these solutions can be used to explore special sectors in ABJM theory
in the same way that the BMN limit can be used to explore a sector of ${\cal N}=4$ super Yang-Mills theory.

An important difference with the BMN case  is that
in that case the limit corresponds  to a Penrose limit of the $\AdS_5\times S^5$ space, where  string theory becomes solvable, allowing for an explicit
comparison between field theory and string theory results.
In the present case, because the M2 branes are extended, it is meaningless to ask what is the geometry seen by the generic null configurations; in particular,
it cannot be obtained as a Penrose limit.
In addition, in the generic case the configurations are non-perturbative from the viewpoint of string theory.
Nevertheless, it is possible that a study of small  fluctuations around these configurations could unveil an interesting sector of the quantum spectrum on $\AdS_4$ and thence
of ABJM ${\cal N}=6$  Chern-Simons theory.

Finally, in section 8, we have revisited the supersymmetric giant graviton solutions found in \cite{japos} representing giant tori and spiky M2 branes.
We re-derived the supersymmetric conditions in cylindrical coordinates, which turn out to be highly convenient
to investigate the solutions in different regimes. This has unveiled a number of interesting supersymmetric uncompact M2 brane objects,
including a cylinder, a hyperboloid and  the giant diabolo, that extend up to the boundary of $\AdS_4$.
They should correspond to deformations ABJM by adding extra degrees of freedom (this is similar to the addition of ``flavor" D7 branes to a D3 brane system).

\section*{Acknowledgements}

We thank Jaume Gomis, Ki-Myeong Lee, Juan Maldacena, David Mateos and especially Paul Townsend for  useful discussions. ARL and JGR
acknowledge support by the research grant ARGEN2007-012.
JGR thanks Universidad de La Plata for hospitality during the course of this work. He also acknowledges support by
research grants MCYT FPA 2007-66665. The work of JLC and JGR has been supported in part by CUR Generalitat de
Catalunya under project 2009SGR502.
ARL would like to thank the Deparment ECM of the U. Barcelona and the Abdus Salam ICTP HECAP Section for hospitality.

\renewcommand{\thesection}{\Alph{section}}
\setcounter{section}{0}

\section{Useful relations}\label{appA}
In the main text we have defined the completely antisymmetric quantities,
\begin{equation}
\begin{aligned}
\beta_{\alpha\beta}^{i j} & = \beta_\alpha^{i} \beta_\beta^{j} - \beta_\alpha^{j} \beta_\beta^{i}\ , \\
\beta^{ijk} & = \beta_0^{i}\; \beta_{12}^{jk} + \beta_0^{j}\; \beta_{12}^{ki} + \beta_0^{k} \beta_{12}^{ij}\ .
\end{aligned}
\end{equation}
In terms of them we have,
\begin{subequations}
\begin{align}
h_c^{00}&=  \left(\frac{R}{2}\right)^{\!\!4}\;\sum_{i<j}\mu_i{}^2\,\mu_j{}^2\; \left(\beta_{12}^{ij}\right)^2\ ,\\
h_c^{11}&= \left(\frac{R}{2}\right)^{\!\!4}\;\left[\sum_{i<j}\mu_i{}^2\,\mu_j{}^2\; \left(\beta_{20}^{ij}\right)^2
-\omega_0{}^2\;\beta_{i,2}\;\beta^i_2\right]\ ,\\
h_c^{22}&= \left(\frac{R}{2}\right)^{\!\!4}\;\left[\sum_{i<j}\mu_i{}^2\,\mu_j{}^2\; \left(\beta_{01}^{ij}\right)^2
-\omega_0{}^2\;\beta_{i,1}\;\beta^i_1\right]\ , \\
h_c^{01}&= \left(\frac{R}{2}\right)^{\!\!4}\;\sum_{i<j}\mu_i{}^2\,\mu_j{}^2\; \beta_{12}^{ij}\;\beta_{20}^{ij}\ ,\\
h_c^{02}&= \left(\frac{R}{2}\right)^{\!\!4}\;\sum_{i<j}\mu_i{}^2\,\mu_j{}^2\; \beta_{12}^{ij}\;\beta_{01}^{ij}\ ,\\
h_c^{12}&= \left(\frac{R}{2}\right)^{\!\!4}\;\left[\sum_{i<j}\mu_i{}^2\,\mu_j{}^2\; \beta_{20}^{ij}\;\beta_{01}^{ij}
-\omega_0{}^2\;\beta_{i,1}\;\beta^i_2\right]\ .
\end{align}
\end{subequations}

For the solution with parameters given in (\ref{betita})  we obtain the following non-zero coefficients,
\begin{equation}
\begin{aligned}
\beta^{12}_{01}&= -4 \omega_2\ , &
\beta^{14}_{01}&= -4 \omega_4\ , &
\beta^{24}_{01}&= -4 m\;\omega_4\cr
\beta^{23}_{02}&= 4 \omega_2\ , &
\beta^{24}_{02}&= 4 n\;\omega_2\ , &
\beta^{34}_{02}&= -4 \omega_4\cr
\beta^{13}_{12}&= 4\ , &
\beta^{14}_{12}&= 4 n\ , &
\beta^{23}_{12}&= 4 m\ , &
\beta^{24}_{12}&= 4 m\;n \ .
\end{aligned}
\end{equation}
and
\begin{align}
\beta^{123}& = -8\;\omega_2\ , &
\beta^{124}&= -8\;n\;\omega_2\ , &
\beta^{134}& = 8\;\omega_4\ , &
\beta^{234}&= 8\;m\;\omega_4\ .
\end{align}
%
These expressions are used in sect.~\ref{8thsusysol}.

\medskip

Finally, we quote the general formula for  the relation between angular momentum and energy  for our  family
of solutions (\ref{sngral}):
\begin{multline}
\frac{1}{R\,E}\;\sum_{i=1}^4 |J_i| = \frac{1}{|3\,z-1|^\frac{1}{2}\,|z-z_0|^\frac{1}{2}}\;\left(
\frac{|z^2 - z_2\,z + (\mu_1{}^2 +\mu_2{}^2)\,z_0|^\frac{1}{2}}{|2\,z - \mu_1{}^2 -\mu_2{}^2|^\frac{1}{2}}
\Big(\mu_1\,|z - \mu_1{}^2|^\frac{1}{2} + \mu_2\,|z - \mu_2{}^2|^\frac{1}{2} \Big)\right.
\\ +
\left. \frac{|z^2 - z_4\,z + (\mu_3{}^2 +\mu_4{}^2)\,z_0|^\frac{1}{2}}{|2\,z - \mu_3{}^2 -\mu_4{}^2|^\frac{1}{2}}
\Big(\mu_3\,|z - \mu_3{}^2|^\frac{1}{2} + \mu_4\,|z - \mu_4{}^2|^\frac{1}{2}\Big)\right)
\end{multline}
From this general expression one can see that for supersymmetric solutions with $z=0$ the r.h.s. is equal to $1$,
giving rise to the BPS  expression (\ref{EJ}).
An interesting question is if there are special values of $\mu_i$ and  $z$
for which the r.h.s. is also equal to 1, hence giving rise to the same BPS expression. This would hint on
special supersymmetric configurations.

\renewcommand{\thesection}{\Alph{section}}

\section{Supersymmetry of collapsed membranes: effective string approach}\label{susystringappA}

As explained in Section 6.3,  in the  case of the collapsed membrane the $\Gamma_\kappa $ of the M2 brane
is singular and cannot be used.
It seems more appropriate  to  study the supersymmetry of this collapsed M2 brane configuration by demanding
the supersymmetry condition under the ``reduced" $\kappa$-symmetry matrix associated with a string-like configuration.
\footnote{We thank J. Maldacena for a discussion on this point.}
 In what follows we will show that this approach correctly reproduces the number of preserved supersymmetries
 obtained in section 6.3 from the supersymmetry algebra. Our results will not rely on the value of the effective
 string tension (classically the string is tensionless, since, on-shell, the world-sheet is null).

\medskip

We consider the following ``reduced''  $\kappa$-symmetry matrix, appropriate for string-like configurations,
\begin{equation}
\Gamma_\kappa = \frac{1}{\sqrt{-g}} \ \dot X^ \mu {X'}^ \nu \Gamma_{\mu\nu}\ ,
\end{equation}
where
\be
 \Gamma_{\mu\nu} = \frac12 \, [\Gamma_\mu ,\ \Gamma_\nu ] \ ,\qquad \Gamma_\kappa^2 =1\ .
 \ee
%
A short computation yields,
\begin{equation}
\Gamma_\kappa = \frac{R^2}{4\, \sqrt{-g}}\;\left( \sum_{i<j}\, \mu_i\, \mu_j\, \beta^{ij}\, \gamma_{i+6}\,\gamma_{j+6}
+ \omega_0\, \sum_i\, \mu_i\, \beta^i_1\, \gamma_0\,\gamma_{i+6} \right) \ ,
\end{equation}
where $\;\beta^{ij} \equiv \beta^{ij}_{01} = \beta^i_0\;\beta^j_1 - \beta^j_0\;\beta^i_1$.
With  these ingredients, and using the relations
\begin{subequations}
\begin{align}
\mathcal M^{-1}\, \gamma_{i+6}\, \mathcal M &=  M_t^{-2}\;\Gamma_0\;
\Big( -\sum_i\;\mu_i\;O_i - \sum_{j\neq i}\;\mu_j\, \gamma_{i+6}\, \gamma_{j+6} \; O_j\;M_i{}^2\;M_j{}^2 \Big) \ ,\\
\mathcal M^{-1}\, \gamma_{i+6}\, \gamma_{j+6}\, \mathcal M &= \gamma_{i+6}\, \gamma_{j+6} \; M_i{}^2\;M_j{}^2 \ ,
\end{align}
\end{subequations}
the supersymmetry condition~(\ref{susycond}) can be written as
\begin{multline}\label{susycollapsed}
\pm\left(   \sum_i\,\mu_i{}^2 m_i{}^2 - \sum_{i<j}  \mu_i{}^2 \mu_j{}^2 \left(\frac{2\,\omega_i}{\omega_0}\;m_j -
\frac{2 \omega_j}{\omega_0} m_i\right)^2\right)^{\!\!1/2} \epsilon_0\\
=\left[ -\sum_i\mu_i{}^2 m_i O_i + \sum_{i<j} \mu_i \mu_j \gamma_{i+6}\, \gamma_{j+6}\, M_i{}^2\;M_j{}^2
\left(m_j\;O_i -m_i\;O_j - \frac{2\,\omega_i}{\omega_0}\;m_j +\frac{2\,\omega_j}{\omega_0}\;m_i\right)\right]\epsilon_0,
\end{multline}
where
\begin{equation}\label{expij}
M_i{}^2\,M_j{}^2 = \exp \left(- \frac12\hat\gamma \gamma_0\,(\beta^i_a\,O_i +\beta^j_a\,O_j) \,
\sigma^a \right) \ .
\end{equation}
This equation is the starting point for analyzing the different possibilities, taking into account that the
$\sigma^a $-dependence on the second term of the r.h.s. must drop out.

%



Let us first consider the generic case  where all $\mu_i$'s are non-vanishing.
We find a  solution by canceling all four $\sigma^a $-dependent terms in the
r.h.s of~(\ref{susycollapsed}) and leaving at least two non-vanishing windings, $m_1, m_2\neq 0$. We have
\be
O_i\; \epsilon_0 =\eta_i\;\epsilon_0\ , \qquad \eta_i{}^2=1\ , \qquad
\frac{2\,\omega^i}{\omega_0} = \eta_i - a\,m_i\ ,
\label{susy42}
\ee
together with the constraint,
\be\label{consign4}
\sum_{i=1}^4\,\mu_i{}^2\;\eta_i\;m_i = \mp\,\left|\sum_{i=1}^4\,\mu_i{}^2\;\eta_i\;m_i\right|\ ,
\ee
where the $\mp $ signs correspond to the $\pm $ signs of  (\ref{susycollapsed}).
Note that one possible solution of the constraint is that the sum in (\ref{consign4}) vanishes. This is indeed the case as seen
from the membrane equations of motion (see  (\ref{consgauge})), although it is not implied by the supersymmetry conditions.
Due to the relation $\prod_{i=1}^4 O_i = -1$ (which fixes $\eta_4 = -\eta_1\,\eta_2\,\eta_3$),
we see that the solution  preserves $1/8$ of the supersymmetries. With no loss of generality we can set $\eta_{1,2,3}=1$, $\eta_4=-1$,
as the signs of $\eta_{i}$ can be reversed by a coordinate transformation
$\xi^{i}\to -\xi^{i}$.
Furthermore, the solution can be rewritten as,
\be
\xi^i = \frac{1}{2}\;\omega_0\;\sigma^0 + m_i\;\sigma'^1\ ,\qquad \sigma'^1 \equiv\sigma^1 -
\frac{a}{2}\,\omega_0\;\sigma^0\ ,\qquad i=1,...,4\ ,
\ee
showing that the string rotates and winds with $m_i$ in each of the four planes.
This also shows that the parameter $a$ is gauged away after the change of coordinate $\sigma\rightarrow\sigma'^1$.


\medskip


Now consider the case where there is a non-trivial embedding in three planes ($(12),(34),(56)$),
i.e.   $\mu_4= 0\;;\; \mu_1{}^2 + \mu_2{}^2 +\mu_3{}^2 = 1$.
A solution is obtained  by cancelling three $\sigma^a $-dependent terms in the r.h.s of~(\ref{susycollapsed}). One needs  at least two non-vanishing
$m_i$, i.e $m_i\neq 0$, $i=1,2$. Now
\be
O_i\; \epsilon_0 =\eta_i\;\epsilon_0\ , \qquad \eta_i{}^2=1\ , \qquad
\frac{2\,\omega^i}{\omega_0} = \eta_i - a\,m_i\ , \qquad i=1,2,3\ ,
\label{susy31}
\ee
together with the constraint (\ref{consign4}).
From (\ref{susy31}) we see that the solution preserves $1/8$ of the supersymmetries.

\medskip

Finally, we consider the two-plane case $\mu_3 = \mu_4 = 0\;;\; \mu_1{}^2 + \mu_2{}^2 = 1$ (thus we take the $(12)$ and $(34)$-planes).
We demand that the $\sigma^a$-dependent term vanishes, leaving at least  two non-vanishing winding numbers, $m_1, m_2\neq 0$.
We need to impose,
\be
O_i\; \epsilon_0 =\eta_i\;\epsilon_0\ ,\quad \eta_i{}^2=1\ ,\qquad \frac{2\,\omega^i}{\omega_0} = \eta_i - a\,m_i\ ,\qquad i=1,2\ ,
\label{susy21}
\ee
together with the constraint (\ref{consign4}).
In view of~(\ref{susy21}), the solution preserves 1/4 of the supersymmetries.
Thus, in all cases,  the number of unbroken supersymmetries obtained from the effective string approach agrees with the results derived in section 6.3 from the superalgebra.


\section{Energy and angular momenta of solutions of section 8}\label{EJdiabolo}

The solution~(\ref{sndiabolo}) is characterized by the energy, some winding numbers and  five angular momenta,
four of which associated with rotations around the $\xi^i$ directions and another one associated with rotations around $\varphi$.
These quantities can be computed directly from the Born-Infeld action
by differentiating with respect to the parameter that governs translations along the corresponding directions.
{}For this, it is convenient to introduce a parameter $\omega_0 $ in
 $t =\sigma^0\to  \omega_0\,\sigma^0$. Later we will set $\omega_0 = 1$ to return to our original solution.

{}For the ansatz~(\ref{ansatzdiabolo}),  the action becomes
\begin{equation}
\begin{aligned}
S&= - \frac12 T_2 \pi  R^{3} \int\!\!\dd z \bigg\{  \bigg[ \bigg(1+{\rho'}^2-{f'}^2\Big)
\left( \omega_0^2 f^2 \alpha_1^2 \rho^2  -  (\alpha_1 s_0-\alpha_0 s_1)^2 \rho^2 + s_1^2  \omega_0^2 f^2 \right) \bigg]^{1/2}
\\ &\qquad\qquad
 + \omega_0 \rho (z\rho'- \rho) \bigg\} \ ,
\end{aligned}
\end{equation}
where we have used ~(\ref{conds}) and  $|m_i |=1$. 
The energy  and the five angular momenta 
can then be obtained
from $E = \left.\frac{\dd S}{\dd\omega_0}\right|_{\omega_0=1}$, $J_i ={\mu_i}^2 \frac{\dd S}{\dd s_0}$ and
$J_\varphi = \frac{\dd S}{\dd\alpha_0}$, leading to the following expressions:
\bea
E &=& \frac{\pi}{2}\; |w| \;T_2\;R^2\;\int\,dz\;\left\{
\frac{\sqrt{1 + \rho'^2-f'^2}\;f^2\; (\rho^2 + b^2)}{\sqrt{f^2\;(\rho^2 + b^2)- (b-1)^2\;\rho^2}} +
\rho\;(z\;\rho' - \rho)\right\}\cr
J_i &=& -\frac{\pi}{2}\; |w| \;T_2\;R^3\;(b-1)\;\mu_i{}^2\eta_1\;e_i\;e_4\;
\int\,dz\;\frac{\sqrt{1 + \rho'^2-f'^2}\; \rho^2}{\sqrt{f^2\;(\rho^2 + b^2)- (b-1)^2\;\rho^2}}\cr
J_\varphi &=& -\frac{\pi}{4}\; |w| \;T_2\;R^3\;b\;(b-1)\;\eta_2\;
\int\,dz\;\frac{\sqrt{1 + \rho'^2-f'^2}\; \rho^2}{\sqrt{f^2\;(\rho^2 + b^2)- (b-1)^2\;\rho^2}}
\eea
where we have substituted  the parameters by their values in (\ref{sndiabolo}), i.e. $\alpha_1= w$ and 
 \begin{equation}
 \begin{aligned}
 s_0 & =  \frac12 \eta_1 (1-b)\ , & s_1 & = - \frac12 \eta_1 \eta_2 \, b \, w  \,  , \\
 s_2 & =0  \ , & \alpha_0 & = 0 \ .
 \end{aligned}
 \end{equation}
We recall that the signs $e_i$, $\eta_{1,2}$ can be chosen as $e_1=e_2=e_3=1$, $e_4=-1$, $\eta_1=\eta_2=1$.
 
Using the specific form of the  solution (\ref{sndiabolo}),
\begin{equation}
z^2  = r_0^2 \rho^c -1 -\rho^2 \ , \qquad f^2 = r_0^2 \rho^c \ , \qquad c\equiv {2b\over b-1}\ ,
\end{equation}
the formulas for the angular momenta take the simple form
\begin{subequations}
\begin{align}
%
J_i & = -\pi |w| \, T_2    R^{3} \mu_i^2  \eta_1  e_ie_4 \int\!\!\dd \rho\ \frac {\rho}{\sqrt{r_0^2 \rho^c -1 -\rho^2} }\ , \\
J_\varphi & =  \frac{\pi}{2}\, |w| \, T_2   R^{3} b\, \eta_2 \  \int\!\!\dd \rho\  \frac {\rho}{\sqrt{r_0^2 \rho^c -1 -\rho^2}}
\ .
\end{align}
\end{subequations}
Note that the change of integration variable from $z$ to $\rho$ introduces  a factor of 2 which accounts for both positive and negative  $z$ integration regions.
Note the simple relations
\begin{equation}
\frac{J_i}{J_\varphi} = - 2\eta_1 \eta_2 e_ie_4 \mu_i^2 \frac1b\ ,\qquad \sum_i |J_i|=  \frac {2}{ |b|}\ |J_\varphi | \ .
\end{equation}
%


\end{document}